\documentclass[a4paper,11pt]{article}
\pdfoutput=1

\usepackage{jcappub}
\usepackage[T1]{fontenc}
\usepackage{soul}
\usepackage{comment}
\usepackage{caption}
\usepackage{subcaption}
\usepackage{xcolor}

\title{Investigating the Origin of CMB Large-Scale Features Using LiteBIRD and CMB-S4}

\author[1]{Catherine Petretti,}
\author[2,3]{Matteo Braglia,}
\author[1]{Xingang Chen,}
\author[3,4,5]{Dhiraj Kumar Hazra,}
\author[6]{Sonia Paban}
\affiliation[1]{Institute for Theory and Computation, Harvard-Smithsonian Center for Astrophysics, 60 Garden Street, Cambridge, MA 02138, USA}
\affiliation[2]{Center for Cosmology and Particle Physics, New York University, 726 Broadway, New York,
NY 10003, USA}
\affiliation[3]{INAF/OAS Bologna, via Gobetti 101, I-40129 Bologna, Italy}
\affiliation[4]{The Institute of Mathematical Sciences, CIT Campus, Chennai 600113, India}
\affiliation[5]{Homi Bhabha National Institute, Training School Complex, Anushakti Nagar, Mumbai
400085, India}
\affiliation[6]{Department of Physics, Harvard University, 17 Oxford St, Cambridge, MA 02138, USA}

\emailAdd{catherine.petretti@cfa.harvard.edu}
\emailAdd{mb9289@nyu.edu}
\emailAdd{xingang.chen@cfa.harvard.edu}
\emailAdd{dhiraj@imsc.res.in}
\emailAdd{sppaban@fas.harvard.edu}

\abstract{Several missions following Planck are currently under development, which will provide high-precision measurements of the Cosmic Microwave Background (CMB) anisotropies. Specifically, measurements of the $E$ modes will become nearly limited by cosmic variance, which, especially when considering the sharpness of the $E$-mode transfer functions, may allow for the ability to detect deviations from the concordance model in the CMB data. We investigate the capability of upcoming missions to scrutinize models that have been proposed to address large-scale anomalies observed in the temperature spectra from WMAP and Planck. To this purpose, we consider four benchmarks that modify the CMB angular power spectra at large scales: models producing suppression, a dip, and amplification in the primordial scalar power spectrum, as well as a beyond-$\Lambda$CDM prescription of dark energy. Our analysis shows that large-scale measurements from LiteBIRD will be able to distinguish between various types of primordial and late-time models that predict modifications to the angular spectra at these scales. Moreover, if these deviations from the standard cosmological model are determined to be systematic and do not reflect the true universe model, future experiments could potentially dismiss these features as statistical fluctuations. We also show that additional measurements from CMB-S4 can impose more stringent constraints by probing correlated signals that these models predict at smaller scales ($\ell\gtrsim 100$).
A byproduct of our analysis is that a recently proposed ``Dark Dimension'' scenario, featuring power amplification at large scales, is strongly bound by current data, pushing the deviation from the standard model to unobservable scales. Overall, our results demonstrate that future CMB measurements can provide valuable insights into large-scale anomalies that are present in the current CMB data.}

\makeatletter
\gdef\@fpheader{}
\makeatother

\begin{document}
\maketitle
\flushbottom

\section{Introduction}\label{sec:intro}

Observations from the Planck satellite currently provide the most precise maps of the CMB temperature and $E$-mode polarization anisotropies \citep{Planck:2018nkj}. These measurements are consistent with a nearly scale-invariant primordial scalar power spectrum with scalar amplitude $A_s = 3.045 \pm 0.016$ and scalar spectral tilt $n_s = 0.9649 \pm 0.0044$, measured at the pivot scale $k_*=0.05~{\rm Mpc}^{-1}$ \citep{Planck:2018jri}. Furthermore, the primordial gravitational wave background predicted by inflation models \cite{Grishchuk:1974ny,Starobinsky:1979ty,Rubakov:1982df}, which is expected to be embedded in the CMB $B$-mode polarization anisotropies on large scales ($\ell\lesssim 100$) \cite{Seljak:1996ti,Seljak:1996gy,Kamionkowski:1996zd}, is constrained in terms of the tensor-to-scalar ratio $r$ to be $r<0.036$ at the pivot scale $k_*=0.05~{\rm Mpc}^{-1}$ at 95\% confidence by the BICEP/Keck collaboration \cite{BICEP:2021xfz} (see also~\cite{Campeti:2022vom,Tristram:2021tvh,Galloni:2022mok,Paoletti:2022anb} for recent reanalyses of these constraints). A key goal of upcoming CMB missions is to probe primordial tensor modes with high enough precision to detect primordial gravitational waves, since, within the vanilla inflationary models, such a detection would provide a direct probe of inflation as well as a direct measurement of its energy scale.

The simplest inflation models are single-field slow-roll (SR) models, in which a scalar field, called the inflaton, rolls down a nearly flat potential \cite{Starobinsky:1980te, Guth:1980zm, Linde:1981mu, Albrecht:1982wi}. While some SR models predict $r$ to be within the realm of observational sensitivities for future experiments, conclusive predictions cannot be made due to unknown physics at very high energies. Furthermore, small-field models, in which the inflaton traverses a sub-Planckian distance, predict a very small tensor-to-scalar ratio $r$ \cite{Lyth:1996im}. It is therefore useful to consider models beyond SR that have predictions we can hope to observe with future experiments.

Primordial features are examples of departures in the potential or kinetic term from SR that induce strongly scale-dependent components in the primordial density perturbations. Besides revealing the shape of the inflaton potential (e.g.~\cite{Starobinsky:1992ts, Adams:1997de,Adams:2001vc,Chen:2008wn,Flauger:2009ab}), primordial features can momentarily increase the energy scale of inflation, potentially boosting the energy reach \cite{Chen:2022vzh} of this ``cosmological collider'' \cite{Chen:2009zp,Arkani-Hamed:2015bza} by orders of magnitude, allowing for exploration of fundamental physics at even higher energies beyond what can be observed in ground-based collider experiments. Primordial features induced by oscillating massive fields could also be used to probe the background evolutionary history of the primordial universe, providing a possible form of direct evidence for the inflation scenario \cite{Chen:2011zf,Chen:2015lza,Chen:2018cgg,Quintin:2024boj}. (See \cite{Chen:2010xka,Chluba:2015bqa,Slosar:2019gvt,Achucarro:2022qrl} for reviews on primordial features.) While there is no evidence for such features in current CMB observations, there are some notable, although not statistically significant, deviations from the concordance model at several multipoles in the measured angular spectra. In particular, deviations at very large scales in the $TT$ power spectrum have been observed by both WMAP and Planck. These so-called ``low-$\ell$ anomalies'' are manifest in the $TT$ spectrum as a suppression of power at $\ell<10$ and a dip at $\ell\simeq 20$, the significance of the former depending on the Galactic mask~\cite{Gruppuso:2015xqa} and the latter being approximately $2\sigma$ away from the $\Lambda$CDM best fit. These anomalies have been explored as possibly being primordial in origin \cite{Peiris:2003ff, Bean:2008na, Mortonson:2009qv,Hazra:2010ve, Hazra:2014goa, Miranda:2014fwa}. While these anomalies may be explained by statistical fluctuations, especially when noting the large uncertainties at $\ell<10$ due to cosmic variance, future experiments may determine whether these deviations are significant, and they are therefore worth exploring.

A slew of next generation CMB missions following Planck are under development or taking data. Some notable examples include ACT \cite{ACT:2020gnv}, SPT \cite{SPT-3G:2024atg}, LiteBIRD \cite{LiteBIRD:2022cnt}, Simons Observatory \cite{SimonsObservatory:2018koc}, and CMB-S4 \cite{CMB-S4:2016ple}. Since Planck has achieved nearly cosmic-variance limited measurements in the temperature anisotropies up to $\ell\sim 2500$, most of the improvement in precision from these missions will be in polarization. LiteBIRD is planned for launch by the 2030s and will cover satellite observations with a focus on polarization at large scales ($\ell\leq 200$). Simons Observatory and CMB-S4 are both ground-based experiments. Simons Observatory began taking data in mid-2024, probing multipoles up to $\ell= 4000$. CMB-S4 observations are also planned for the 2030s, probing multipoles up to $\ell = 5000$. One of the primary goals of these experiments is to obtain high-precision $B$-mode measurements in order to detect primordial gravitational waves. Since the amplitude of the $B$-mode anisotropies is small compared to that of $E$-mode and temperature anisotropies, these missions require unprecedented precision of $E$-mode measurements, with the $EE$ and $TE$ power spectra projected to being constrained nearly to the level of cosmic variance. (Planck measurements have already achieved precision near the level of cosmic variance in the $TT$ power spectrum.) Since models designed to explain the low-$\ell$ anomalies in the $TT$ spectrum also predict correlated features in the $EE$ and $TE$ spectra, future experiments will provide a more accurate test of these models.

In this paper, we test the ability of future CMB experiments to constrain or detect cosmological models beyond $\Lambda$CDM that modify the CMB power spectra at large scales with a particular emphasis on primordial feature models. Current candidates to explain the low-$\ell$ anomalies include such primordial feature models, which modify the shape of the scalar primordial power spectrum from inflation, as well as models that alter the late-time evolution of the universe, affecting the integrated Sachs-Wolfe (ISW) effect. These models also predict deviations in the CMB bispectra, as well as in the matter power spectrum and bispectrum from Large Scale Structure (LSS), which could help constrain the models further. As a representative sample, we consider three popular models with distinct large-scale effects in the CMB power spectrum: (1) a primordial feature introduced as a step in the inflationary potential that can produce either a bump or a dip at large scales, used to explain the $\ell\simeq 20$ dip in $TT$; (2) a second primordial feature model introducing a kink in the potential, which can lead to either large-scale suppression or amplification, used to explain the $\ell<10$ suppression in $TT$; and (3) a model of dark energy parameterized by the equation of state $w$ of the dark energy fluid and the speed of sound $c_s^2$ of its perturbations, modifying the late-time ISW effect and also producing either large-scale suppression or amplification. For completeness, we also consider a fourth model, characterized by a micron-sized fifth spacetime dimension that predicts only amplification of the scalar primordial power spectrum. 
We use the fully Bayesian pipeline introduced in \cite{Braglia:2021ckn} and later improved in \cite{Braglia:2022ftm}. We first compare the models against real Planck data, providing constraints on their parameters, identifying best-fit candidates, and assessing their observational evidence compared to the standard cosmological model. We then perform a Bayesian forecast analysis for the $TT$, $EE$, and $TE$ power spectra using the projected sensitivities to temperature and $E$ modes from future experiments. While the main driver of our analysis is large-scale measurements from LiteBIRD, most models aiming to fit the low-$\ell$ anomalies produce modifications to the anisotropy spectra at smaller scales as well. Therefore, we complement our forecast with projected smaller-scale measurements. While such high-quality anisotropy measurements will soon be available from the Simons Observatory, our purpose is to set the most optimistic prospects by including small-scale data, so we provide projected forecasts from CMB-S4. As a final test of LiteBIRD's capabilities, we ignore the short-scale effects of the primordial feature models and only focus on the suppression and dip anomalies at large scales, which we represent with non-physical, diagnostic templates that produce no small-scale modifications to the primordial power spectrum. We repeat our data and forecast analyses on these templates. We aim to answer three major questions with our analysis: (1) If the true model of the universe is featureless (i.e., consistent with the concordance model), can we hope to rule out large-scale features as statistical fluctuations? (2) Conversely, if the true model contains some kind of large-scale feature, can we detect such a feature? (This question has also been studied for the step model in \cite{Miranda:2014fwa}.) (3) In the case of (2), can we further discriminate between different types of features with future observations?

Our paper is outlined as follows: we begin by describing the four models used in our analysis in Sec. \ref{sec:models}. We first constrain the models using the existing Planck data. We present an overview of the methodology used for our analysis in Sec. \ref{sec:methods}, and the results are summarized in Sec. \ref{sec:bf}. In Sec. \ref{sec:forecast}, we present the results of our forecast analyses for LiteBIRD and CMB-S4. We also present data and forecast analyses of non-physical, diagnostic templates for the low-$\ell$ anomalies in Sec. \ref{sec:templates}. We conclude and discuss implications of our results in Sec. \ref{sec:concl}.

\section{The models}\label{sec:models}

We begin by presenting an overview of the models used in this study. We consider three types of large-scale deviations from the concordance model: a dip, suppression, and an amplification  in the $TT$ power spectrum. Our goal is to investigate if future experiments are able to detect such deviations and discriminate between them.

Our baseline model, which we refer to as the ``featureless'' model, is the vanilla $\Lambda$CDM model supplemented by an initial stage of SR inflation. This model, which is described by only six cosmological parameters, provides a very good fit to cosmological data~\cite{Planck:2018vyg}. To model the initial stage of SR inflation, we consider the minimal scenario where it is driven by a scalar field $\phi$ with a canonical kinetic term and a minimal coupling to gravity, whose action is given by
\begin{equation}
 S = \int {\rm d}^4 x \sqrt{-g}    \left[-\frac{1}{2} (\partial\phi)^2-V(\phi)\right].
 \end{equation}
 In particular, we choose to work with a small-field hilltop toy model specified by the potential
\begin{equation}
V_0(\phi)= V_{\rm inf} \left(1- \frac{1}{2} C_\phi \phi^2 \right).
    \label{eq:potential}
\end{equation}
This model is convenient because it allows us to easily relate the free parameters of the model to the scalar power spectrum amplitude $A_s$ and tilt $n_s$ as well as to the tensor-to-scalar ratio $r$, all defined at a given scale with the Hubble crossing condition $k_0=a_0H_0$

\begin{equation}
    A_s = \frac{V_{\rm inf}}{24\pi^2 \epsilon_0}, \quad n_s-1 =-2C_\phi,\quad r=16\epsilon_0=8\,C_\phi^2 \phi_0^2. \label{eq:eff_params1}
\end{equation}
From these equations, we see that $\epsilon_0$ controls the amplitude of the tensor-to-scalar ratio $r$. Since our analysis in this paper does not make use of $B$-mode data, we follow previous works~\cite{Braglia:2021ckn,Braglia:2021sun,Braglia:2021rej,Braglia:2022ftm} and fix $\epsilon_0=10^{-7}$ ($r\approx 10^{-6}$) in order not to produce observable $B$ modes.

{\bf  Dip at Large Scales.} We first consider a sharp primordial feature introduced as a bump or dip in the primordial power spectrum. Such a feature can be generated by a ``step'' in the inflaton potential. 
The step model was originally motivated by a supergravity model in which several stages of spontaneous symmetry breaking during inflation trigger a series of steps in the inflaton potential
\cite{Adams:1997de, Adams:2001vc}, and could have other origins as well. After WMAP, such a model is used to fit the $\ell\simeq 20$ dip observed in the $TT$ angular power spectrum \cite{Peiris:2003ff, Bean:2008na, Mortonson:2009qv, Dvorkin:2009ne, Hazra:2010ve, Dvorkin:2011ui, Hazra:2014goa, Miranda:2014fwa}.
Phenomenologically, the inflaton potential of this model may be written in the following form \cite{Adams:2001vc},
 
\begin{equation}
    V(\phi) = V_0(\phi)\left[1+\delta_{\rm step}(\phi) \right], \quad \delta_{\rm step}(\phi)= \frac{A}{2}\left[1+\tanh\left(\frac{\phi_0-\phi}{\Delta}\right)\right] ~,
    \label{eq:step}
\end{equation}
where $A$ is the height of the step, $\phi_0$ is the feature location, and $\Delta$ is the feature width. The density perturbation in the power spectrum at the feature location can be described as either a dip if $A>0$ (a downward step) or a bump if $A<0$ (an upward step). 
This model introduces three model parameters: $A$, $\phi_0$, and $\Delta$. We relate these model parameters to the following three ``effective parameters,'' which are more intuitive and more closely related to observables: the maximum deviation in the primordial power spectrum from the featureless spectrum, $\Delta \mathcal{P}/\mathcal{P}_0$; the number of $e$-folds from the beginning of inflation until the inflaton encounters the sharp feature, $N_0$; and the feature duration in terms of number of $e$-folds, $\Delta N$. The relations between the model and effective parameters are derived in Appendix \ref{app:eff_pars}.

{\bf Large-Scale Primordial Suppression.} Next, we consider a primordial feature model that results in large-scale suppression by introducing a ``kink'' in the inflaton potential.
The kink model was originally engineered \cite{Starobinsky:1992ts}  as an attempt to reconcile a non-$\Lambda$CDM (matter-only) model with the observed galaxy correlation functions. We will use this model phenomenologically despite the original motivation.
The potential of the kink model may be written in the following form,
\begin{equation}
    V(\phi)=V_0(\phi)\left[1+\delta_{\rm kink}(\phi)\right], \quad \delta_{\rm kink}=  \frac{A}{2}(\phi_0-\phi)\left[ 1+\tanh \left( \frac{\phi_0-\phi}{\Delta}\right) \right]~,
    \label{eq:kink}
\end{equation}
where $A$ is the initial slope of the potential, $\phi_0$ is the feature location, and $\Delta$ is the width of the kink. This feature is characterized by a sharp change in the shape of the potential, which is of the form $V\sim V_0(\phi)[1+A(\phi_0-\phi)]$ for $\phi\ll\phi_0$ and $V\sim V_0$ for $\phi\gg\phi_0$. This model can result in either suppression in the power spectrum if $A>0$ (a negative initial slope) or amplification if $A<0$ (a positive initial slope). This model is therefore motivated by fitting the $\ell<10$ suppression in the $TT$ angular power spectrum. The same model parameters introduced in the step model also describe the kink model, and these are similarly related to the effective parameters $\Delta \mathcal{P}/\mathcal{P}_0$, $N_0$, and $\Delta N$. We note that other models exist that predict a similar suppression of power that could fit the low-$\ell$ suppression observed in the temperature data. For example, double inflation models where inflation consists of two stages driven by two kinetically coupled scalar fields produces a feature signal similar to that of the kink model~\cite{Braglia:2020fms}. Wiggly Whipped Inflation (WWI)~\cite{Hazra:2014goa} combines a step-like feature at intermediate scales with large-scale suppression~\cite{Hazra:2014jka} arising from phase transitions during inflation\footnote{A forecast with the WWI model using a Cosmic ORigins Explorer (CORE) like survey was done in~\cite{Hazra:2017joc}.}. Other single field models such as ``punctuated inflation''~\cite{Jain:2008dw,Jain:2009pm}, fast roll initial conditions on the inflaton~\cite{Contaldi:2003zv,Hazra:2014jka,Cicoli:2014bja,Hergt:2018ksk}, inflationary models with non-zero spatial curvature~\cite{Hergt:2022fxk,Letey:2022hdp}, or pre-inflationary physics~\cite{Gruppuso:2015xqa,Gruppuso:2015zia,Gruppuso:2017nap} also produce suppression, although more drastic than that considered here.

{\bf Large-Scale Late-Time Suppression.} In addition to considering primordial models, we also investigate whether future CMB experiments can distinguish such features from large-scale suppression produced by post-inflationary effects. We focus on contributions to the ISW effect, in which CMB photons are redshifted as they traverse large regions of high gravitational potential due to the expansion of the universe.
This effect can be observed as an enhancement in the $TT$ power spectrum on large scales ($\ell \lesssim 10$) \cite{Sachs:1967er}.
The ISW contribution to the $TT$ spectrum depends on the gravitational potential $\Phi$ along the line of sight \cite{Sachs:1967er,Hu:1994uz}, 
which is related to the total comoving density perturbation $\delta \rho_{\rm tot}$ by the Poisson equation \begin{equation}
    k^2 \Phi = -4\pi Ga^2 \delta \rho_{\rm tot}.
\end{equation}
Only matter and dark energy contribute to the ISW effect, and thus $\delta \rho_{\rm tot} = \delta\rho_m + \delta \rho_{\rm de}$. For a general non-interacting fluid with a constant equation of state $w\equiv p/\rho$, the fractional density perturbations $\delta\equiv \delta\rho/\rho$ evolve according to
\begin{equation} 
 \delta'+3\mathcal{H}(c_s^2-w)(\delta+3\mathcal{H}(1+w)v/k)+(1+w)kv=-3(1+w)h' \label{eq:pert} 
\end{equation}
and
\begin{equation}
    v'+\mathcal{H}(1-3c_s^2)v+kA=k c_s^2\delta/(1+w), \label{eq:velocity}
\end{equation}
where $\mathcal{H}$ is the conformal Hubble parameter, $c_{s}^2\equiv \delta p/\delta \rho$ is the sound speed of the perturbations comoving with the fluid, $v$ is the velocity, $A$ is the acceleration, and $h'\equiv (\delta a/a)'$ is the local expansion rate \cite{Weller:2003hw}, where the derivatives are taken with respect to conformal time. These equations assume negligible anisotropic stress, which is the case for matter perturbations and simple dark energy models. Also note that a varying $w$ leads to additional ISW contributions not expressed above.

Notice from Eq. (\ref{eq:pert}) that $w=-1$ has the solution $\delta=0$, meaning that a cosmological constant has no fluid density perturbations. If $w\neq -1$, however, perturbations are expected. For quintessence models, in which the dark energy fluid is related to a scalar field with a simple canonical kinetic term $\mathcal{L}_{de}=\pm \frac{1}{2}(\partial_\mu \varphi)^2-V(\varphi)$, $w$ can differ from $-1$ and in general can also evolve with time \cite{Carroll:2003st}. Here, the perturbations evolve with sound speed equal to the speed of light $c_s^2=1$. In other models, the Lagrangian can be more complicated and can therefore account for a generic sound speed \cite{Armendariz-Picon:2000nqq}. Furthermore, both $c_s^2$ and $w$ can, in general, evolve with time. We consider such models with generic values of the comoving sound speed $c_s^2$ and dark energy equation of state $w$, both of which we will assume to be time independent for simplicity, following the prescription given in \cite{Weller:2003hw}. We therefore refer to this model as the ``$w-c_s^2$'' model.

To understand the impact of $w$ and $c_s^2$ on the $TT$ power spectrum, let us first ignore the effect of the perturbations and focus only on the equation of state. If $w<-1$, the dark energy density $\rho_{\rm de}$ increases with time, resulting in a more dramatic decrease of $\Phi$ with time. This results in CMB photons having a higher overall redshift, and the ISW contribution is enhanced compared to the cosmological constant case. The opposite effect occurs if $w>-1$, and the ISW contribution is suppressed. However, these results also depend on the value of $c_s^2$, and in fact, the amount of amplification or suppression due to the value of $w$ alone can be entirely canceled by the effects of perturbations.

In general, a small $c_s^2$ results in more clustering of dark energy perturbations, and conversely, the perturbations are smoothed out quickly if $c_s^2\sim 1$. If $w<-1$ and $c_s^2$ is small, the increased $\rho_{\rm de}$ with time combined with the clustering results in a more enhanced ISW contribution compared with a larger sound speed. If $c_s^2$ is larger, the dark energy perturbations smooth out quickly, resulting in less enhancement in the $TT$ power spectrum. As $c_s^2$ approaches the speed of light, the perturbations smooth out so quickly that their impact effectively cancels the ISW enhancement due to the equation of state. On the other hand, if $w>-1$ and $c_s^2$ is small, even though the dark energy perturbations cluster more effectively, because the expansion rate is smaller than the case where $w\geq -1$, the effect of these perturbations does not contribute as significantly to $\Phi$. This results in a more suppressed ISW effect compared to the case where $c_s^2$ is larger. If the sound speed is larger, the perturbations smooth out more quickly, which allows the perturbations to contribute more to $\Phi$. This counteracts the effect of a decreasing $\rho_{\rm de}$ and results in less suppression in the $TT$ power spectrum. Again, as $c_s^2$ approaches unity, the effect of the perturbations essentially cancels the effect of $w>-1$. 

Finally, while our analysis focuses on the CMB and the ability of this model to fit the low-$\ell$ anomalies, it is important to note that models with a time-evolving equation of state also significantly impact the background evolution of the universe and the late-time clustering of matter. Consequently, they are strongly constrained by other datasets, such as supernovae and LSS data. In particular, the equation-of-state parameters are degenerate with other background parameters such as the matter, dark energy, and curvature densities. Late-time observations help in constraining the equation of state further by breaking certain degeneracies. However, since we analyze the constraints and forecasts for CMB-only observations, we do not consider additional datasets. Therefore, the constraints on this model presented here should be considered conservative and are expected to become more stringent with the inclusion of these additional datasets.

{\bf Large-Scale Amplification.} Lastly, we consider a class of models that results only in amplification on large scales to investigate the ability of CMB measurements to test such features. Motivated by the Swampland program \cite{Vafa:2005ui}, a fifth spacetime dimension of size $\sim 1-10~\mu$m, called the ``dark dimension,'' has been proposed as a solution to the cosmological hierarchy problem \cite{Montero:2022prj, Gonzalo:2022jac, Obied:2023clp, Agmon:2022thq, Vafa:2024fpx}. To account for the relatively large size of this compact dimension compared to the 5D Planck scale (around $10^9 ~\rm{GeV}$),  recent research \cite{Anchordoqui:2023etp} has suggested a period of inflation in 5D. This period would end with the scale of the compact dimension being locked at a minimum.  This inflation shares the benefits of 4D inflation  but predicts a power spectrum that violates near scale invariance at large scales while maintaining scale invariance for shorter scales \cite{Antoniadis:2023sya, Anchordoqui:2023etp}. Notably, the cut-off scale is associated with five-dimensional physical wavelengths of around $1~\mu$m. It is precisely this behavior that makes it a suitable candidate for the analysis presented in this paper.

Following \cite{Antoniadis:2023sya}, we model the primordial power spectrum for 5D inflation as
\begin{equation}
\begin{aligned}
    \mathcal{P}(k) =& A_s \left\{\left(\frac{k}{k_*}\right)^{2\delta-5\epsilon}\left[\coth x+x~{\rm csch}^2x \right] +\frac{\epsilon}{72} \left(\frac{k}{k_*}\right)^{-3\epsilon} \right. \\
    &\left.\times\left[2x^3{\rm csch}^4x+15\coth x+x~{\rm csch}^2x\left(4x^2\coth^2x+12x\coth x+15\right) \right]\right\}, \label{eq:ddim}
\end{aligned}
\end{equation}
where $\epsilon$ and $\delta$ are respectively the first and second SR parameters. We have defined the variable $x\equiv \pi kR_0$, where $R_0$ is the compact dimension radius at the beginning of inflation with the scale factor normalization chosen in \cite{Anchordoqui:2023etp}. Since the initial value of $R_0$ is not fixed   in \cite{Antoniadis:2023sya}, we convert it into a parameter to be determined by our analysis. We relate $R_0$ to the effective parameter $N_{\rm DD}-N_*$ via 
\begin{equation}
    R_0 \equiv \frac{1}{\pi k_{\rm DD}} = \frac{1}{\pi k_* e^{N_{\rm DD}-N_*}}.
\end{equation}
We use the convention $k_*\equiv 0.05~{\rm Mpc}^{-1}$, which exits the horizon at $N_* \equiv N_{\rm end}-50$ $e$-folds from the start of inflation ($N_{\rm start}\equiv 0$), and where $N_{\rm DD}$ is the number of $e$-folds from the start of inflation at which the dark dimension scale $k_{\rm DD}$ exits the horizon.

The large-scale and small-scale limits of Eq. (\ref{eq:ddim}) to leading order are
\begin{equation}
    \mathcal{P}(k) \underset{k\ll k_{\rm DD}}{\simeq} 2A_s\frac{k_{\rm DD}}{k}\left(\frac{k}{k_*}\right)^{2\delta-5\epsilon} \quad {\rm and} \quad
    \mathcal{P}(k) \underset{k\gg k_{\rm DD}}{\simeq} A_s \left(\frac{k}{k_*}\right)^{2\delta-5\epsilon},
\end{equation}
respectively. This demonstrates that $\mathcal{P}(k)$ is nearly scale invariant in the small-scale limit with the spectral tilt given by
\begin{equation}
    n_s-1 = 2\delta - 5 \epsilon . \label{eq:ns}
\end{equation}
In the large-scale limit, the spectral tilt is given by $n_s-1 = 2\delta-5\epsilon-1$, which, for $\delta,\,\epsilon\ll1$ implies that the power spectrum increases towards smaller scales with $n_s\sim 0$. In our template, we use $A_s\equiv \mathcal{P}(k_*)$, $n_s$, and $N_{\rm DD}-N_*$ as effective parameters for the dark dimension model. Also, we note that other models exist which predict large-scale amplification, and that we use the dark dimension model as a recent example. 

\begin{figure}[ht]
    \centering
    \includegraphics[width=0.75\textwidth]{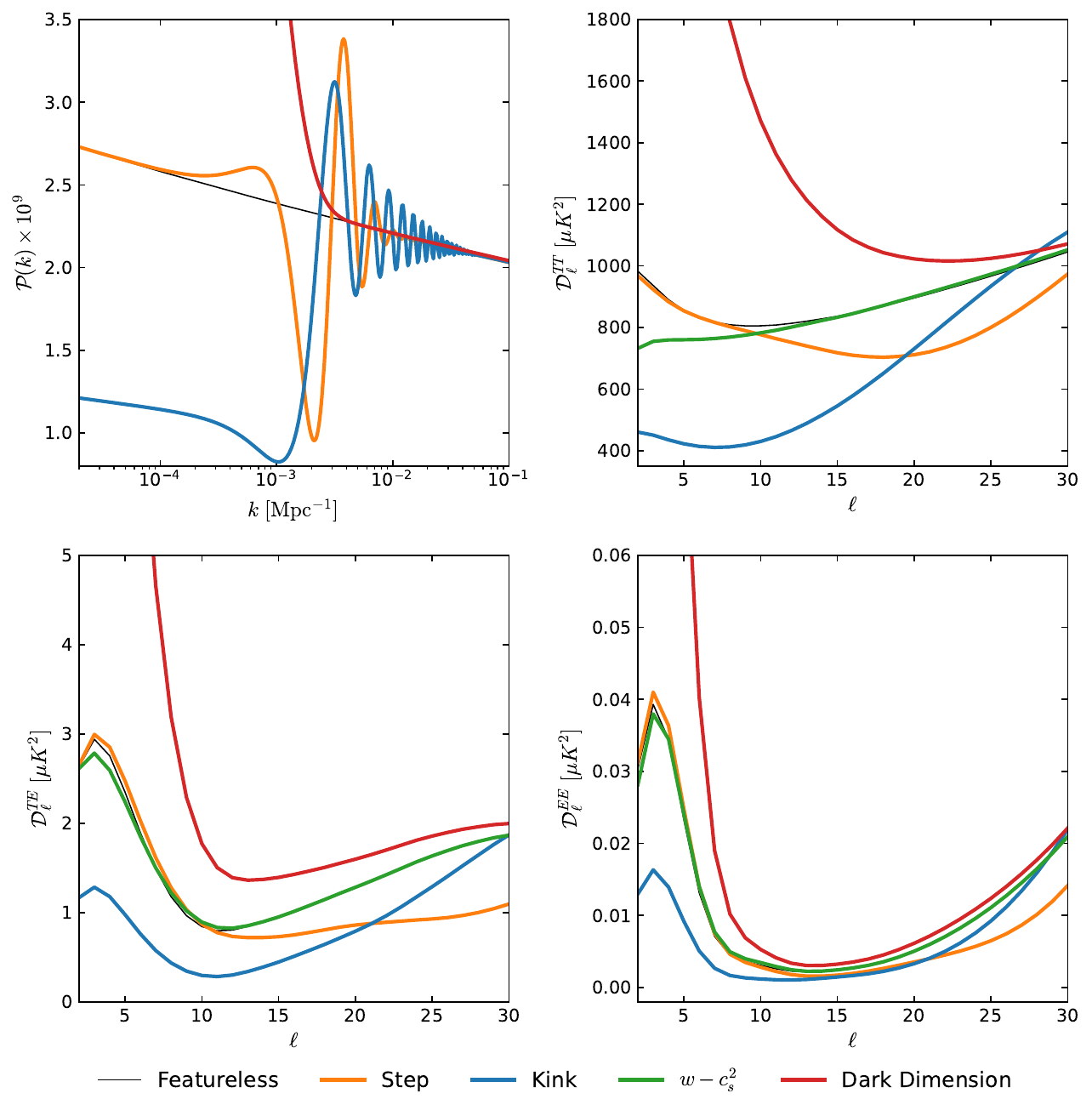}
    \caption{Examples of large-scale effects in the primordial, $TT$, $TE$, and $EE$ power spectra [from left to right] for each of the models considered in this paper. For these examples, the step model parameters are $\ln (10^{10}A_s)=3.14$, $n_s=0.966$, $N_0-N_*=-4.0$, $\Delta N=0.5$, and $\Delta \mathcal{P}/\mathcal{P}_0=-0.6$. The kink model parameters are the same as those for the step model except for $\Delta N=0.1$ and $\Delta \mathcal{P}/\mathcal{P}_0=-0.5$. The featureless and $w-c_s^2$ models were calculated using the same parameters for the kink model but with $\Delta \mathcal{P}/\mathcal{P}_0=0$. The additional $w-c_s^2$ model parameters are $w=-0.5$ and $c_s^2=0.01$, with $H_0=53.0~{\rm km}~{\rm s}^{-1}~{\rm Mpc}^{-1}$. The dark dimension model parameters are $\ln (10^{10}A_s)=3.04$, $n_s=0.966$, and $N_{\rm DD}-N_*=-4.0$ with $A_s$ and $n_s$ calculated at $k_*=0.05~{\rm Mpc^{-1}}$. These example parameters were chosen to demonstrate the general characteristics of the low-$\ell$ anomalies in $TT$, and observable large-scale amplification for the dark dimension model.}
    \label{fig:dl_examples}
\end{figure}

Examples of the primordial power spectra as well as the temperature and polarization angular power spectra at large scales are shown in Figure \ref{fig:dl_examples} for each model. The step model roughly matches the featureless model until $\ell\sim 20$, where the model exhibits a dip which will be used to match the large-scale dip in the $TT$ power spectrum. In contrast, the kink model exhibits large-scale suppression for $\ell\lesssim 30$. Note that the sign of $\Delta \mathcal{P}/\mathcal{P}_0$ can be changed to result in amplification for the kink model or a bump for the step model. Also, while both the kink model and the $w-c_s^2$ model can result is suppression or amplification on large scales, the $w-c_s^2$ model only affects $\ell \lesssim 10$ while oscillations in the kink model extend to smaller scales. Furthermore, the ISW effect has no impact on $E$ modes, so the $w-c_s^2$ model has no signature in the $EE$ spectrum. The dark dimension model exhibits amplification in the temperature and polarization spectra, but unlike the kink and $w-c_s^2$ models, in which the parameters can be changed to result in either amplification or suppression, this model can only result in an amplification of power.

\section{Methodology} \label{sec:methods}
In this section, we present an overview of the methods used in our data analysis. For additional details, see \cite{Braglia:2021ckn}.

We first constrain the effective parameters using the latest CMB temperature and $E$-mode polarization data from the Planck mission \cite{Planck:2018nkj}. For the multipole range $\ell=2-29$, we use the \texttt{commander\_dx12\_v3\_2\_29} (lowT) likelihood for the temperature anisotropies and the \texttt{simall\_100x143\_offlike5\_EE\_Aplanck\_B} (lowE) likelihood for the $E$-mode anisotropies\footnote{The Planck likelihoods are available on the Planck Legacy Archive: \url{https://pla.esac.esa.int/pla/}. To compute the likelihoods, we use the Planck Likelihood code \texttt{clik 16.0}, which available at \url{https://github.com/benabed/clik}.}. For small angular scales $\ell=30-2500$, we use the \texttt{2020 CamSpec v12.5}\footnote{\url{https://people.ast.cam.ac.uk/~stg20/camspec/}} (EG20) likelihood \citep{Efstathiou:2019mdh}. Because of the wide range of possible effective parameters that could be chosen given the CMB data, especially considering the large uncertainties at low $\ell$, we expect multimodal posterior distributions in our samples. Therefore, we use nested sampling to explore the parameter space using \texttt{CosmoChord}, a \texttt{PolyChord}\footnote{\url{https://github.com/PolyChord}} add-on of \texttt{CosmoMC}\footnote{\url{https://cosmologist.info/cosmomc/}} \citep{Lewis:2002ah,Handley:2015fda,Handley:2015vkr}.

\begin{table}[ht]
    \centering
    \begin{tabular}{|c|c|}
        \hline
        Cosmological Parameters & Priors \\ \hline
        $\Omega_{\rm b} h^2$ & $[0.02,0.0265]$ \\ \hline
        $\Omega_{\rm CDM} h^2$ & $[0.1,0.135]$ \\ \hline
        $100 * \theta_s$ & $[1.03,1,05]$ \\ \hline
        $\tau$ & $[0.03,0.08]$ \\
        \hline
    \end{tabular}
    \begin{tabular}{|c|c|c|c|c|}
        \hline
        Effective Parameters & \multicolumn{4}{|c|}{Priors} \\
        \hline
         & Step & Kink & $w-c_s^2$ & Dark Dimension \\ \hline
        $\ln \left( 10^{10} A_s \right)$ & \multicolumn{4}{|c|}{$[2.763,4.375]$} \\ \hline
        $n_s$ & \multicolumn{4}{|c|}{$[0.920,0.996]$} \\ \hline
        $N_0-N_*$ & \multicolumn{2}{|c|}{$[-5.0,-0.5]$} & $-4.0$ & --- \\ \hline
        $\Delta N$ & \multicolumn{2}{|c|}{$[0.05,1.5]$} & $1.0$ & --- \\ \hline
        $\Delta \mathcal{P}/\mathcal{P}_0$ & $[-0.99,0.99]$ & $[-0.75,0.75]$ & $0$ & --- \\ \hline
        $w$ & \multicolumn{2}{|c|}{$-1$} & $[-2.0,0.0]$ & $-1$ \\ \hline
        $\log_{10} c_s^2$ & \multicolumn{2}{|c|}{$0$} & $[-4.0,0.0]$ & $0$ \\ \hline
        $N_{\rm DD}-N_*$ & \multicolumn{3}{|c|}{---} & $[-6.6,-2.1]$ \\ \hline
    \end{tabular}
    \caption{Uniform priors on the cosmological parameters and effective parameters for each model used in our analysis. The priors on nuisance parameters are kept unchanged from their default values. Parameters where only one value is listed are kept fixed at those values.}
    \label{tab:priors}
\end{table}

Our choice of ranges for the priors of the cosmological and effective parameters are listed in Table \ref{tab:priors}. We purposefully choose a prior volume large enough to account for a wide range of possible effects, including the possibility of a featureless spectrum so that the baseline model is also nested into the extended ones. As mentioned, we set the energy scale of inflation to $\epsilon_0\equiv 10^{-7}$. 
For the featureless, step, kink, and $w-c_s^2$ models, we calculate the primordial power spectrum numerically using \texttt{BINGO}~\cite{Hazra:2012yn}. By convention, we set the end of inflation to $N_{\rm end}=68$ and the pivot scale to exit the horizon at $N_*=N_{\rm end}-50=18$ $e$-folds from the onset of inflation. We also define $A_s$ and $n_s$ at the scale $k_0$, which exits the horizon $N_0$ $e$-folds from the start of inflation, corresponding to the feature scale for the step and kink models. For the featureless and $w-c_s^2$ models, we set $\Delta \mathcal{P}/\mathcal{P}_0=0$ so that $N_0-N_*$ and $\Delta N$ are arbitrary, as no features show up in $\mathcal{P}(k)$. We fix them to $N_0-N_*=-4.0$ and $\Delta N=1.0$. For the dark dimension model, the primordial power spectrum is evaluated directly using Eq. (\ref{eq:ddim}). For this model, $A_s$ and $n_s$ are defined at the pivot scale $k_*\equiv 0.05 ~{\rm Mpc}^{-1}$, which exits the horizon at $N_*=N_{\rm end}-50$. The compact dimension scale $k_{\rm DD}$ exits the horizon at $N_{\rm DD}$ $e$-folds from the onset of inflation.\footnote{Note that the dark dimension scale $k_{\rm DD}$ is different from what we call the feature scale $k_0$ for the other primordial models we consider. Specifically, $k_{\rm DD}$ does not correspond to the point where the transition from $n_s\sim 0$ to $n_s\sim 1$ takes place, which one could associate with the feature scale $k_0$. Instead, we note that $k_0\approx 5k_{\rm DD}$, or $N_0-N_*=\ln 5+N_{\rm DD}-N_*$, which arises from the $\coth x + x{\rm csch}^2x$ term in Eq. (\ref{eq:ddim}). We choose our prior range for $N_{\rm DD}-N_*$ such that $N_0-N_*$ in all of our primordial models has the same prior range.}

To determine the improvement in the likelihood of one model over another, we report the Bayes factor for each model, defined as the ratio of evidences between two models. We calculate the Bayes factor as $\ln B=\ln \mathcal{Z}_{\rm model}-\ln \mathcal{Z}_{\rm featureless}$. In our analysis, a negative $\ln B$ means that the model is statistically disfavored compared to the featureless case, and a positive value means that the model is favored over the featureless case. We interpret the Bayes factors using the Jeffreys' scale \cite{jeffreys1961theory}.

Because we are considering a large number of parameters in our analysis, including cosmological and nuisance parameters, our sampling is not efficient at determining the set of best-fit parameter values for the observational data. This is especially true for models that are not very well-constrained by the data, as the best-fit parameters may not be directly correlated with the peaks in the posterior distributions and are therefore not easily inferred from nested sampling. We use the methods explained above to identify regions of high likelihood and narrow the parameter space, and, using these results, we run the minimizer \texttt{BOBYQA} \cite{powell2009bobyqa} on the negative-log likelihood to find the best fit to the data. Note,  however, these these are merely best-fit ``candidates'' because, as we will shortly see, the extended classes of models have not been detected in Planck observations but may be detected in future experiments. We report our total $\chi^2$ to the data as
\begin{equation}
    \chi^2=\chi^2_{\rm EG20}+\chi^2_{\rm lowT}+\chi^2_{\rm lowE}+\chi^2_{\rm prior},
\end{equation}
where $\chi^2_{\rm prior}$ corresponds to Gaussian priors on particular nuisance parameters (see \cite{Braglia:2021ckn}). We calculate the improvement in the best-fit candidates compared to the baseline model as $\Delta \chi^2=\chi^2_{\rm featureless}-\chi^2_{\rm model}$ for each model.

\section{Results of data analysis} \label{sec:bf}

\begin{table}[ht]
    \centering
    \begin{tabular}{|c|c|c|c|c|}
    \hline
     Parameters & Kink & Step & $w-c_s^2$ & Dark Dimension \\ \hline
    $N_0-N_*$ & -3.69 & -4.20 & \multicolumn{2}{|c|}{---} \\ \hline
    $\Delta N$ & 0.355 & 0.321 & \multicolumn{2}{|c|}{---} \\ \hline
    $\Delta \mathcal{P}/\mathcal{P}_0$ & -0.106 & -0.364 & \multicolumn{2}{|c|}{---} \\ \hline
    $w$ & \multicolumn{2}{|c|}{---} & -1.93 & --- \\ \hline
    $\log c_s^2$ & \multicolumn{2}{|c|}{---} & -0.0649 & --- \\ \hline
    $N_{\rm DD}-N_*$ & \multicolumn{3}{|c|}{---} & -6.44 \\ \hline
    \end{tabular} \\

    \begin{tabular}{|c|c|c|c|c|c|}
        \hline
        Model & $\Delta \chi^2_{\rm total}$ & $\Delta \chi^2_{\rm lowT}$ & $\Delta \chi^2_{\rm lowE}$ & $\Delta\chi^2_{\rm EG20}$ & $\Delta\chi^2_{\rm prior}$ \\ \hline
        Kink & 3.479 & 2.889 & -0.030 & 0.532 & 0.0884 \\ \hline
        Step & 5.229 & 4.247 & 0.178 & 0.564 & 0.240 \\ \hline
        $w-c_s^2$ & 2.568 & 0.363 & 0.459 & 1.701 & 0.045 \\ \hline
        Dark Dimension & -0.213 & -0.476 & -0.040 & 0.067 & 0.236 \\ \hline
    \end{tabular}

        \begin{tabular}{|c|c|c|c|c|}
    \hline & Step & Kink & $w-c_s^2$ & Dark Dimension \\ \hline
    $\ln B$ & $-1.79\pm0.35$ & $-1.60\pm 0.35$ & $-0.79\pm 0.35$ & $-1.45\pm 0.35$ \\ \hline
    \end{tabular}
    \caption{[Top] Best-fit candidates to the Planck data for each model. [Center] $\Delta \chi^2$ for the best-fit candidates with respect to the best-fit featureless candidate. [Bottom] Bayes factors for each model with respect to the featureless model from our Planck data analysis.}
    \label{tab:models_bf}
\end{table}

\begin{figure}[ht]
    \centering
     \begin{subfigure}[c]{0.7\textwidth}
         \centering
         \includegraphics[width=\textwidth]{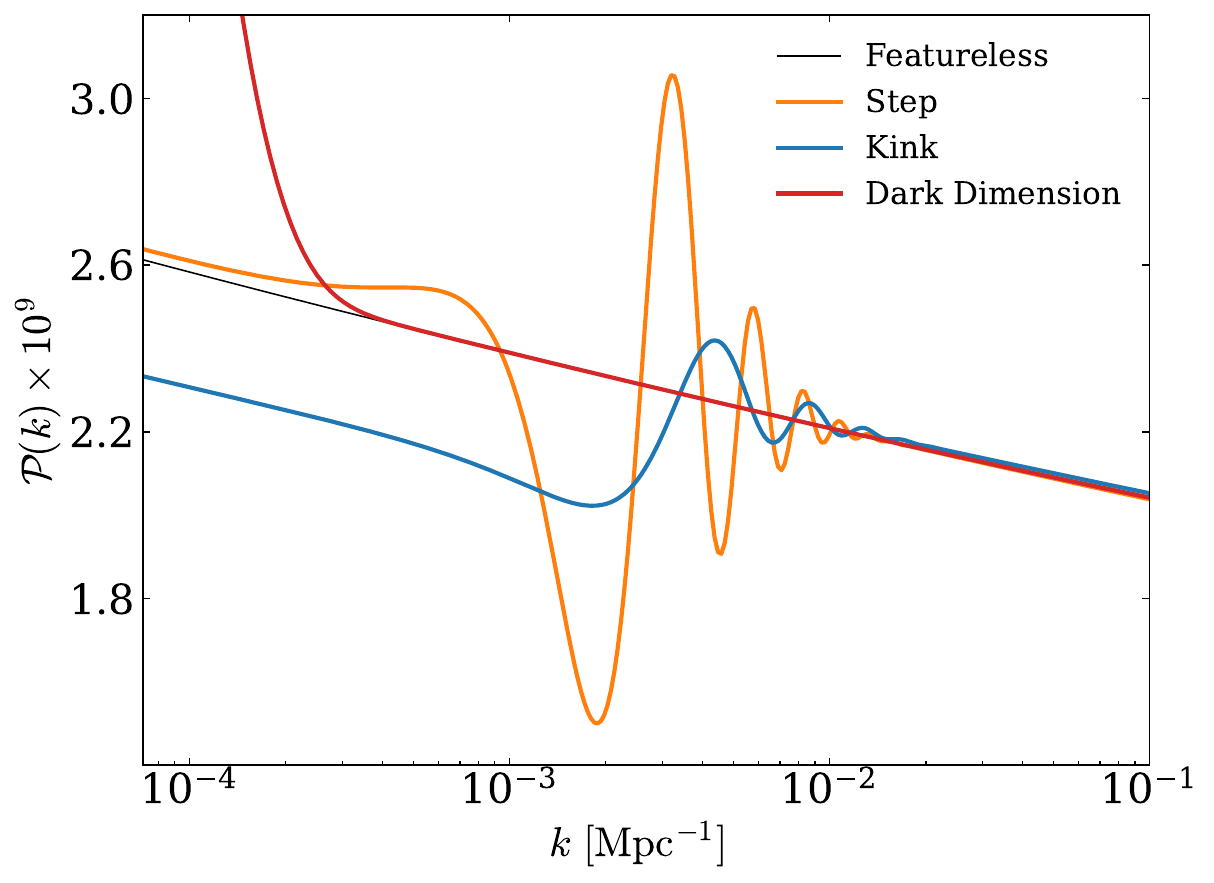}
     \end{subfigure}
     
     \begin{subfigure}[c]{\textwidth}
         \centering
         \includegraphics[width=\textwidth]{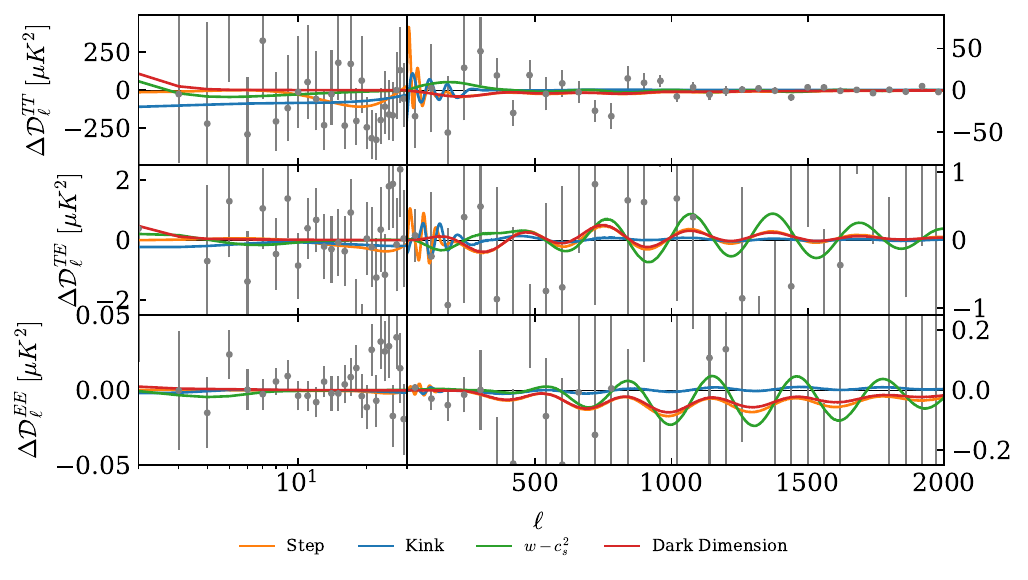}
     \end{subfigure}
     \caption{[Top] Primordial power spectra of the best-fit candidates. The best-fit featureless candidate is represented by the black dashed line.
     Note that the ``best-fit" candidate for the DD model is an artifact from the boundary of priors (see the main texts and appendix for more details).
     [Bottom] CMB power spectrum residuals of the best-fit candidates with respect to the best-fit featureless candidate. The residuals of the Planck measurements with respect to the best-fit featureless candidate are represented by the gray points.  For a better visualization, we bin high-$\ell$ data ($\ell>30$) using a binwidth $\Delta\ell=60$.}
     \label{fig:bf}
\end{figure}
\subsection{Best fit candidates}
Table \ref{tab:models_bf} lists the best-fit values for the parameters for each model as well as the improvement in $\chi^2$ for the models compared to the featureless best-fit candidate. Figure \ref{fig:bf} shows the primordial power spectra and the CMB $TT$, $TE$, and $EE$ power spectrum residuals, calculated with respect to the featureless best-fit candidate. Notice that the dip in the step model candidate aligns with the $\ell\simeq 20$ dip, and the suppression in the kink model candidate fits the $\ell<10$ suppression, as expected. However, the dip in the step model candidate is more extended and shallower than the $\ell\simeq 20$ dip in the data. This is due to the shorter-scale ($30\lesssim \ell\lesssim 250$) oscillations in the step model candidate. Since the data at these scales do not differ significantly from the featureless best fit, the amplitudes of these oscillations, which are also determined by $\mathcal{P}/\mathcal{P}_0$, are constrained and thus limit the amplitude of the $\ell\simeq 20$ dip. The short-scale oscillations may also be limiting the amplitude of the kink model candidate. However, the large-scale suppression is more consistent with the $\ell<10$ suppression exhibited by the data. It is also important to keep in mind that precision in this regime is largely limited by cosmic variance, which limits our ability to estimate the amplitude of the suppression in $TT$.

The best-fit candidate for the $w-c_s^2$ model does not exhibit as much deviation from $\Lambda$CDM at large scales as the kink model best fit. The reason for this is that the Planck data favor an equation of state of $w<-1$ \cite{Planck:2018vyg}, and thus the best-fit parameter for $w$ in our analysis is close to $-2$. However, $w<-1$ results in an amplified ISW effect, unless $c_s^2\sim 1$, as perturbations with a high sound speed will cancel this effect. The resulting $TT$ power spectrum is similar to the featureless spectrum at large scales. Also, for our best-fit equation of state of $w\sim -2$, the density parameters and $H_0$ will be significantly different from their observed values. Indeed, for the best-fit $w-c_s^2$ candidate, we obtain $\Omega_{\rm m}\approx 0.15$ and $H_0\approx 99~{\rm km}~{\rm s}^{-1}~{\rm Mpc}^{-1}$. The best-fit candidate is therefore able to capture the acoustic peaks in the CMB power spectra, so most of the contribution to $\chi^2$ comes from the \texttt{EG20} likelihood (i.e., high-$\ell$ multipoles). However, $w\sim -2$ has been ruled out by BAO measurements \cite{eBOSS:2020yzd}, which are consistent with a cosmological constant. Furthermore, the best-fit values for $\Omega_m$ and $H_0$ have been ruled out by BAO and local measurements. Our best-fit candidate for the $w-c_s^2$ model is not a meaningful result, and constraints on such dark energy models must be placed in conjunction with constraints from other data sets. Our best-fit candidate is solely for illustrative purposes to test late-time models predicting large-scale suppression.

Lastly, the ``best-fit" candidate for the dark dimension model approaches to the featureless baseline model. This candidate is not at the location of the minimum $\chi^2$ which for this model is at the limit of the featureless baseline model, but is an artifact of the boundary of the chosen prior range.\footnote{One might wonder why the best fit value for $N_{\rm DD}-N_*$ is not $-6.6$, corresponding to the edge of our chosen prior. Since $N_{DD}-N_*=-6.43$ is nearly indistinguishable from a featureless model within observable scales (accounting for the cosmic variance at low $\ell$), decreasing $N_{DD}-N_*$ further only yields a negligible improvement in $\chi^2$. However, it is worth pointing out that, while the minimizer can get close to a global minimum, if the search is initiated from a different point in parameter space, there can be small differences in the converged values of the $\chi^2$. The tolerance of $\chi^2$ is set to 0.1 in the search. Therefore, this slight difference of $\chi^2$ ($\sim -0.2$) between $N_{\rm DD}-N_*=-6.6$ and $N_{\rm DD}-N_*=-6.43$ can be attributed to differences in the converged points by the minimizer. Nevertheless, for any small value of $N_{DD}-N_*$ which corresponds to a power law on cosmological scales, the best fit is $\Delta \chi^2\sim 0$  with respect to the baseline model. We also tested that further lowering the lower limit of the $N_{DD}-N_*$ prior reduces its ``best-fit" value as expected.} 
This result can be readily understood: the signature of this model within the observable scales provides a worse fit to the data since the angular power spectra from this model exhibit amplification at low multipoles, but the current data prefers the opposite trend.

\subsection{Model comparison}

Table~\ref{tab:models_bf} also lists the Bayes factors for each model with respect to the featureless model from our analysis of the Planck data. Recall that a negative $\ln B$ means that the model is disfavored by the data compared to featureless. From these results, there is substantial evidence to disfavor both the step and kink models compared to the featureless model. Due to the large uncertainties in the data at low $\ell$, namely in the $E$-mode anisotropies as these measurements are not cosmic-variance limited, the improvement in the likelihood for these models is not significant compared to the featureless model. Therefore, the three additional parameters in these models are penalized in the calculation of the Bayes factor, resulting in these models being disfavored by the Planck data.

There is no clear evidence against the $w-c_s^2$ model given the Planck data. Similar to the results for the primordial feature models, it is unclear whether two additional parameters are needed to better fit the data due to the large uncertainties at low $\ell$. However, the primordial feature models exhibit shorter-scale oscillations where the data have been more precisely measured, contributing to the evidence against these models. In contrast, since the $w-c_s^2$ model has no short-scale effects, no such claim can be made.

The dark dimension model is also substantially disfavored by the Planck data. We have discussed the constraints on this model in detail in Appendix~\ref{sec:DD-contour} and plotted the parameter posteriors in Figure.~\ref{fig:ddim_chi}. From our results we can conclude that,  compared to the baseline model, the observable signatures of this model are substantially ruled out by the data.

\section{Forecast analysis} \label{sec:forecast}

Next, we use the best-fit candidates found in Sec. \ref{sec:bf} to perform a forecast analysis for upcoming CMB surveys. For more details on this procedure, see \cite{Braglia:2021rej}. We aim to answer the following: (1) if the true model of the universe is featureless, can we statistically rule out large-scale features, and (2) if the true model exhibits features at large scales, can we detect such features and distinguish them from other models?

We assume one of our best-fit candidates to be the ``fiducial,'' i.e. the ``true'' model of the universe. We consider only the multiple range $\ell=2-2500$, corresponding to the scales probed by Planck. We simulate observations by adding noise to the fiducial according to the noise curves of a given combination of surveys. Note that these noise curves
refer to the designs discussed in \cite{LiteBIRD:2022cnt,CMB-S4:2016ple}, and the designs are expected
to be constantly updated prior launch of the surveys. We use combinations of Planck data with LiteBIRD (PL + LB) and CMB-S4 (PL + LB + S4). We include CMB-S4 in our analysis to place additional constraints at small angular scales, while LiteBIRD will have the most constraining power at large scales. We assume PL to cover $\ell=800-2500$ with 20\% sky coverage, LB to cover $\ell=2-39$ with 80\% sky coverage and $\ell=40-1350$ with 20\% sky coverage, and S4 to cover $\ell=40-2500$ with 40\% sky coverage. These ranges ensure insignificant overlap between the surveys where their signal-to-noise ratios are comparable. The noise in the power spectrum for each survey is calculated as
\begin{equation}
    \mathcal{C}_\ell^{\rm Noise}=\frac{4\pi f_{\rm sky}\sigma^2}{Nb_\ell (\theta)^2},
\end{equation}
where $f_{\rm sky}$ is the fraction of sky coverage for each survey, $\sigma$ is the r.m.s. noise per pixel, $N$ is the number of pixels covering the sky, and $b_{\ell}(\theta)$ is the beam function for the observations, which we assume to be a Gaussian given by
\begin{equation}
    b_\ell(\theta) = \exp\left(-\frac{\ell(\ell+1)\theta^2}{2}\right)
\end{equation}
with width $\theta$. We use an inverted Wishart likelihood following \cite{Hamimeche:2008ai}.

We fit each model to these simulated data using \texttt{CosmoChord}. We keep the same ranges on the priors for the cosmological and effective parameters as our Planck data analysis (see Table \ref{tab:priors}). We calculate the Bayes factor $\ln B = \ln \mathcal{Z}_{\rm fiducial} - \ln \mathcal{Z}_{\rm model}$ to assess the evidence for or against each model compared to the fiducial. A positive $\ln B$ means that the fiducial is favored, and we interpret the values of $\ln B$ using the Jeffreys' scale \cite{jeffreys1961theory}. In the following discussion, we do not present results of forecasts for the dark dimension model since our data analysis has shown that such large-scale amplification is already severely constrained by the Planck data alone.

\subsection{Results assuming a featureless fiducial}

\begin{figure}[ht]
    \centering
     \begin{subfigure}[b]{0.49\textwidth}
         \centering
         \includegraphics[width=\textwidth]{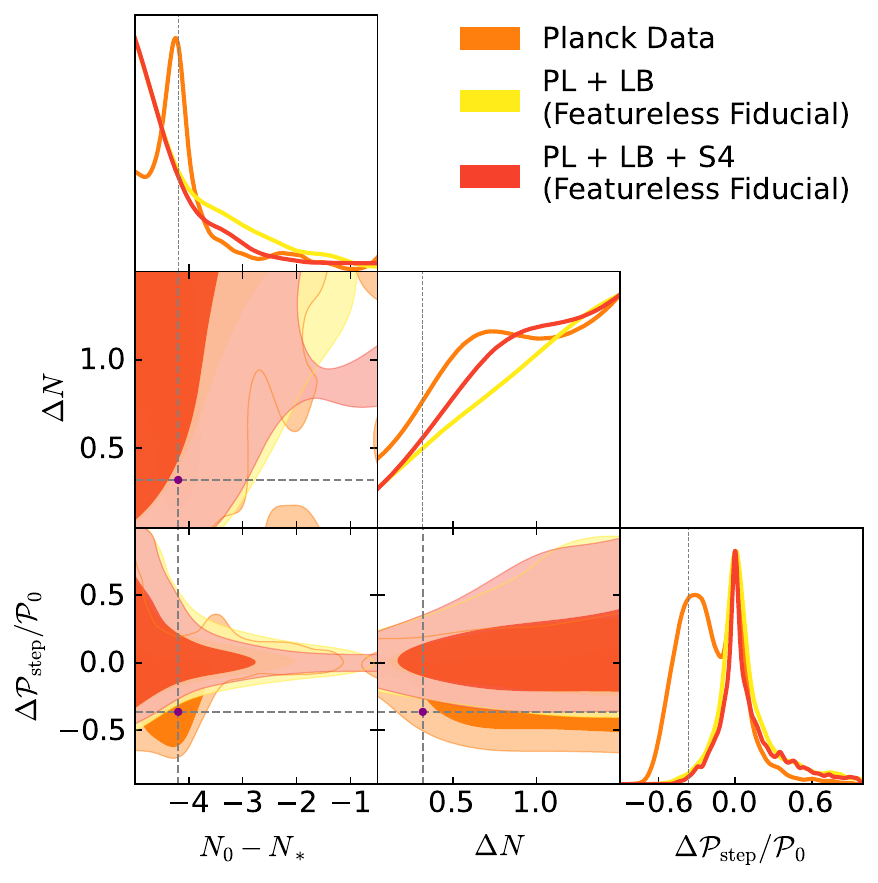}
     \end{subfigure}
     \hfill
     \begin{subfigure}[b]{0.49\textwidth}
         \centering
         \includegraphics[width=\textwidth]{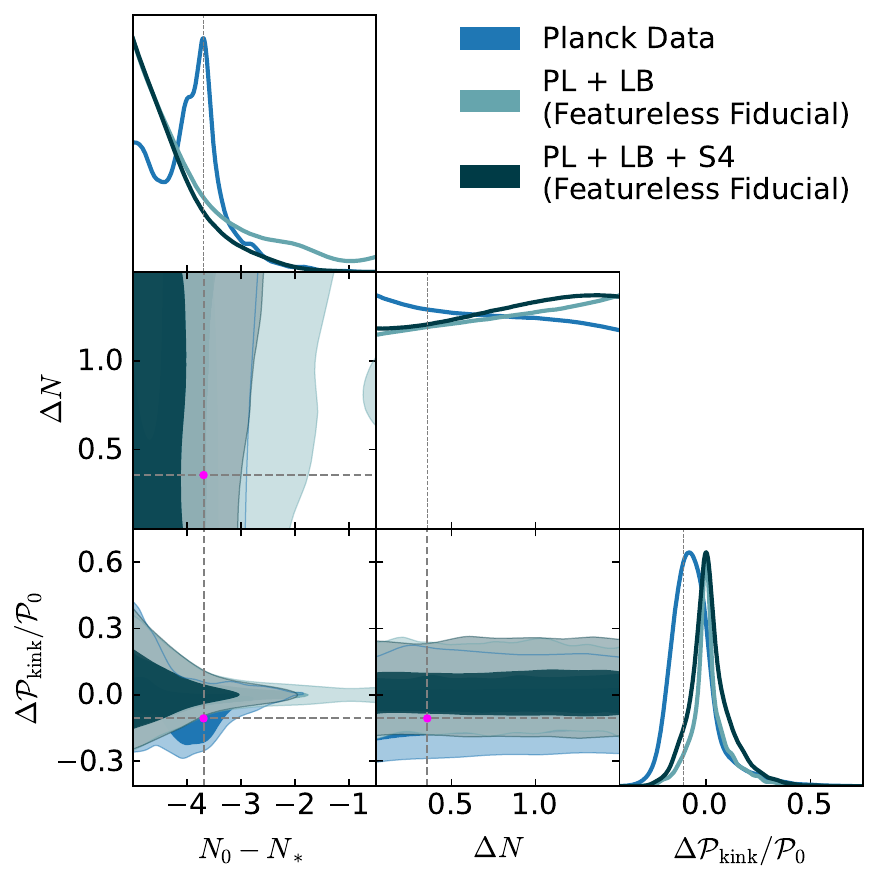}
     \end{subfigure}
     \begin{subfigure}[b]{0.55\textwidth}
         \centering
         \includegraphics[width=\textwidth]{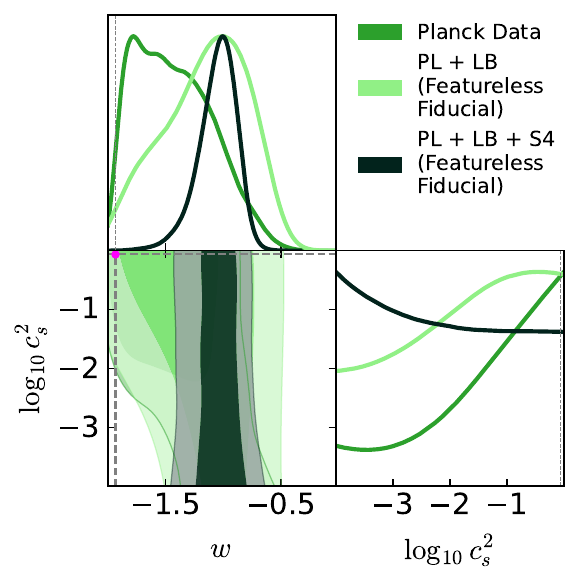}
     \end{subfigure}
     \caption{Constraints on the step [top left], kink [top right], and $w-c_s^2$ [bottom] models assuming a featureless fiducial. We mark the Planck best fit for each model with the small circles.}
     \label{fig:model_to_base}
\end{figure}

We begin by asking whether we are able to rule out large-scale features as statistical fluctuations. To this end, we assume the best-fit featureless candidate as our fiducial and analyze the feature models. Figure \ref{fig:model_to_base} shows the constraints on each model from our Planck data analysis and from the two experimental forecasts (PL + LB and PL + LB + S4). For the primordial feature models, the forecasts more strongly constrain $\Delta\mathcal{P}/\mathcal{P}_0$ to be close to zero. $\Delta N$ is largely unconstrained, which is due to the model construction. The limit $\Delta N \rightarrow 0$ corresponds to infinitely many oscillations in the primordial power spectrum which can consequently overfit the data. Alternatively, the limit $\Delta N \rightarrow \infty$ smooths out the feature. In this limit, the step feature is completely removed and becomes featureless, corresponding to an increased likelihood in the forecasts for large $\Delta N$, while the kink model still maintains the large-scale suppression/amplification. Also, note that small values of $N_0-N_*$ (large scales) are unconstrained. This is because, similar to our Planck analysis of the dark dimension model (see Appendix~\ref{sec:DD-contour}), the feature can be pushed to arbitrarily large scales outside of the observable range and thus resemble a featureless model in observations.

\begin{figure}[ht]
    \centering
    \includegraphics[width=\linewidth]{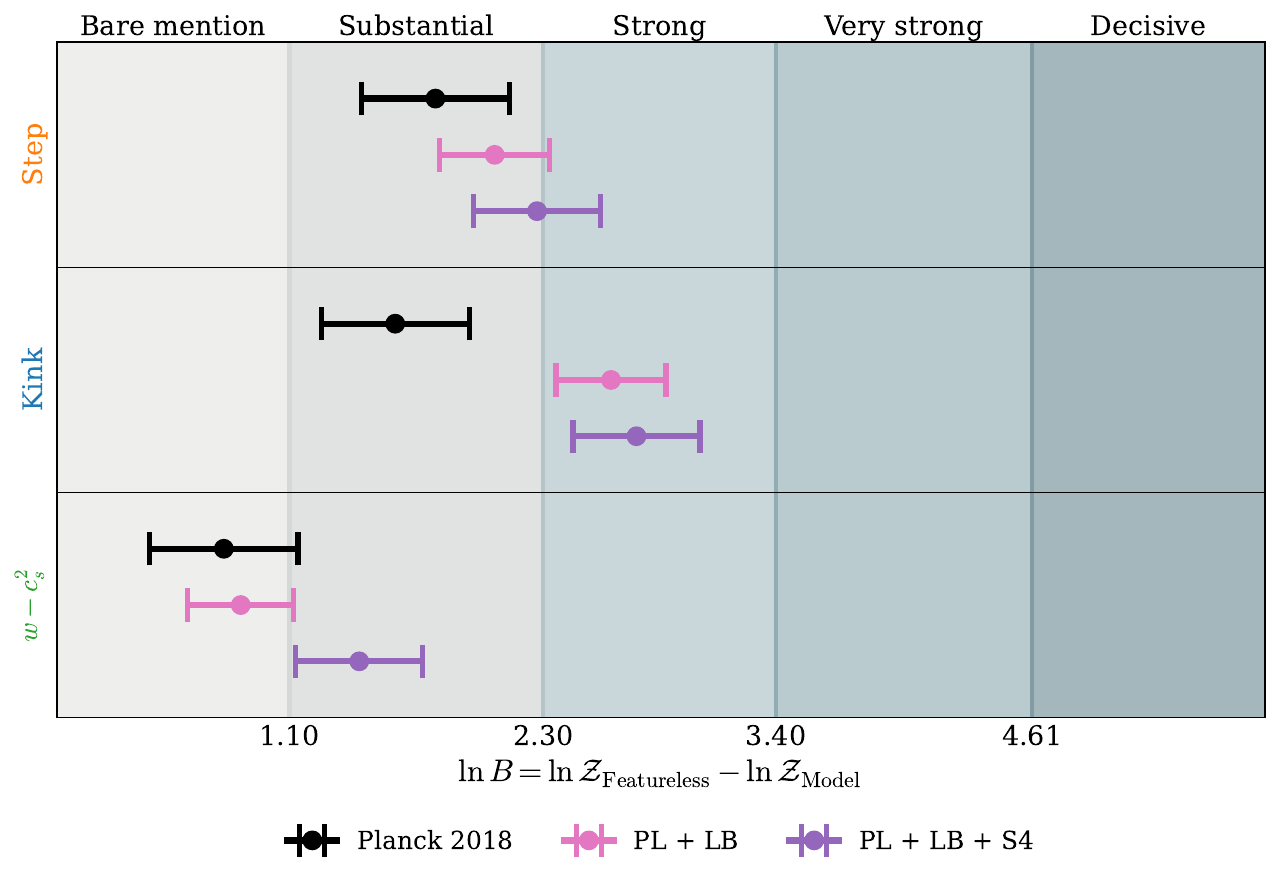}
    \caption{The projected Bayes factors obtained from our data and forecast analyses with respect to the featureless model.}
    \label{fig:model_to_base_bayes}
\end{figure}

For the PL + LB forecast of the $w-c_s^2$ model, a cosmological constant ($w=-1$) is slightly more favored, and the constraint becomes even tighter for the PL + LB + S4 forecast, making the result more consistent with BAO measurements \cite{eBOSS:2020yzd}. This is most likely because the acoustic peaks in the CMB power spectra will be more precisely measured, especially by CMB-S4, and therefore the density parameters and $H_0$ will be better constrained, resulting in a more accurate measurement of $w$. However, since $w=-1$ results in no dark energy perturbations, the value of the $c_s^2$ parameter is arbitrary, resulting in $c_s^2$ being unconstrained. Because of the tighter bound around $w=-1$, this effect on $c_s^2$ is most notable in the PL + LB + S4 forecast.

Figure \ref{fig:model_to_base_bayes} shows the Bayes factors for these results. When adding the simulated measurements from LiteBIRD and CMB-S4, the kink model is strongly disfavored and the step model is substantially-to-strongly disfavored compared to the featureless fiducial. While the Bayes factors for the PL + LB + S4 forecasts between the step and kink models are consistent within $1\sigma$, the mean $\ln B$ for the step model is slightly smaller than that of the kink model. This could be because for the step model, two parameters can be adjusted to recover a featureless model --- either the $\Delta N\rightarrow \infty$ limit or the $\Delta \mathcal{P}/\mathcal{P}_0\rightarrow 0$ limit can give a featureless power spectrum --- while for the kink model, only the $\Delta \mathcal{P}/\mathcal{P}_0\rightarrow 0$ limit results in a featureless spectrum. For the $w-c_s^2$ model, there is a slight increase in the evidence against this model in the PL + LB forecast compared to the Planck analysis, but the evidence is still barely mentionable. When adding CMB-S4, the evidence against the $w-c_s^2$ model becomes substantial. Again, this is likely because CMB-S4 will be able to provide a better constraint on $w$ that is consistent with the $\Lambda$CDM model, making $c_s^2$ an irrelevant parameter and providing more evidence against this model. LSS measurements in conjunction with CMB-S4 would provide even better constraints on $w$.

Note that, since the parameters for each of these models can be adjusted to recover a featureless power spectrum, none of these models can be decisively ruled out, as the amplitude of the feature can be made too small to probe with future CMB experiments. Additionally, for the primordial feature models, the feature location $N_0-N_*$ can be made small enough to push the feature outside of the observable range. Therefore, if the true model of the universe is featureless, large-scale features can be substantially-to-strongly, but not decisively, ruled out by future experiments.

\subsection{Results assuming a fiducial with features}

\begin{figure}[ht]
    \centering    
    \begin{subfigure}[b]{0.49\textwidth}
    \centering
    \includegraphics[width=\textwidth]{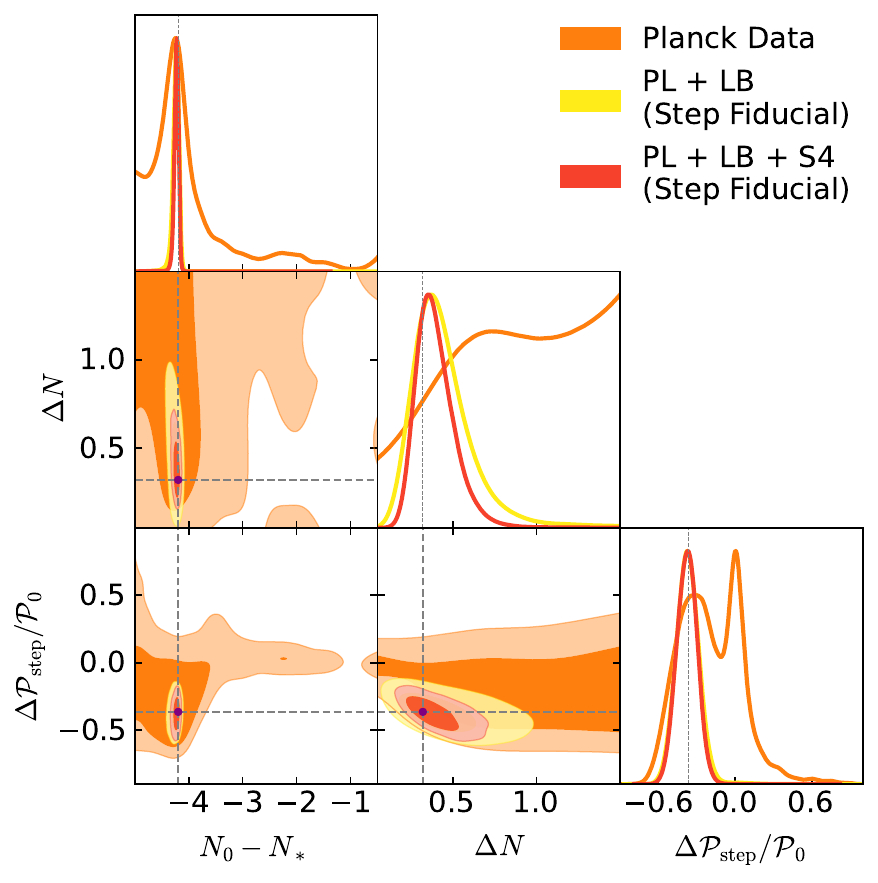}
    \end{subfigure}
    \hfill
    \begin{subfigure}[b]{0.49\textwidth}
    \centering
    \includegraphics[width=\textwidth]{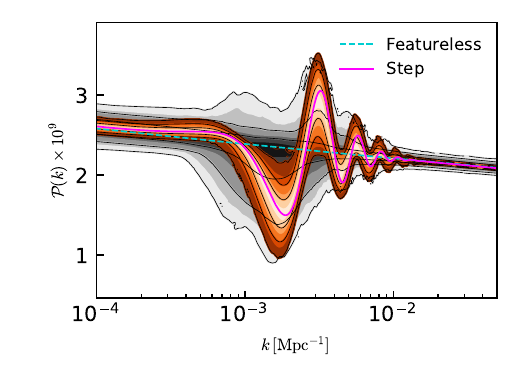}
    \end{subfigure}
    \caption{[Left] Constraints on the effective parameters for the step model from our Planck data analysis and our forecast analyses, assuming a step fiducial. We mark the Planck best fit for the model with small circles. [Right] The reconstructed primordial power spectra for the step model. The reconstructions are shown in gray for the Planck data and in orange for our PL + LB + S4 forecast. The different shades show ranges from 0 (innermost region) to $3\sigma$ (outermost region), and the lines show the $1\sigma$, $2\sigma$, and  $3\sigma$ ranges. The best-fit featureless and step model candidates are plotted in teal and pink, respectively.}
     \label{fig:step_to_step}
\end{figure}

Next, we aim to answer the following: if the true model of the universe exhibits primordial features on large scales, can such features be detected by upcoming experiments? Furthermore, can different types of large-scale features (both primordial and post-inflationary) be distinguished? We assume the best-fit step model candidate to be the fiducial as this model gave the best overall improvement in $\chi^2$ over the featureless best fit (see Table \ref{tab:models_bf}). First, we present constraints on the step model parameters from future experiments in Figure \ref{fig:step_to_step}. Both the PL + LB and PL + LB + S4  forecasts can provide tighter constraints on the effective parameters than Planck alone. The addition of LiteBIRD will yield nearly cosmic-variance limited $E$-mode measurements at the largest scales, providing most of the constraining power for these parameters. Figure~\ref{fig:step_to_step} also shows reconstructions of the primordial power spectrum from the Planck data and the PL + LB + S4 forecast. This plot further demonstrates the tight parameter constraints around the best-fit step candidate from the forecast compared to Planck data alone.

\begin{figure}[ht]
    \centering
     \begin{subfigure}[c]{0.49\textwidth}
         \centering
         \includegraphics[width=\textwidth]{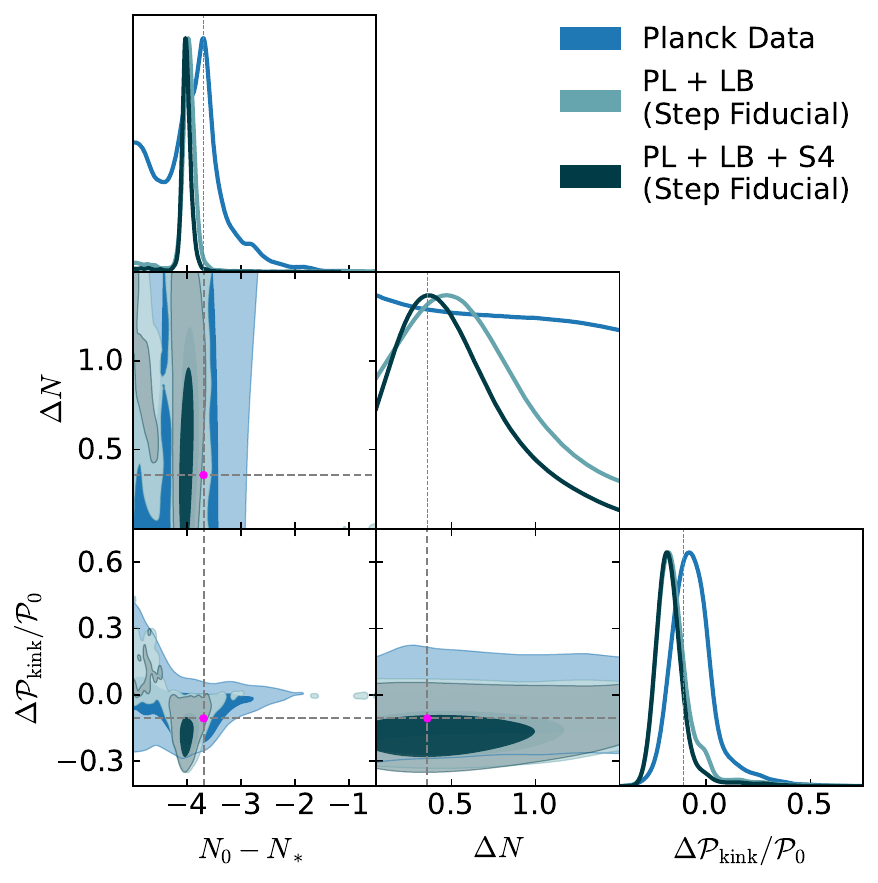}
     \end{subfigure}
     \hfill
     \begin{subfigure}[c]{0.49\textwidth}
         \centering
         \includegraphics[width=\textwidth]{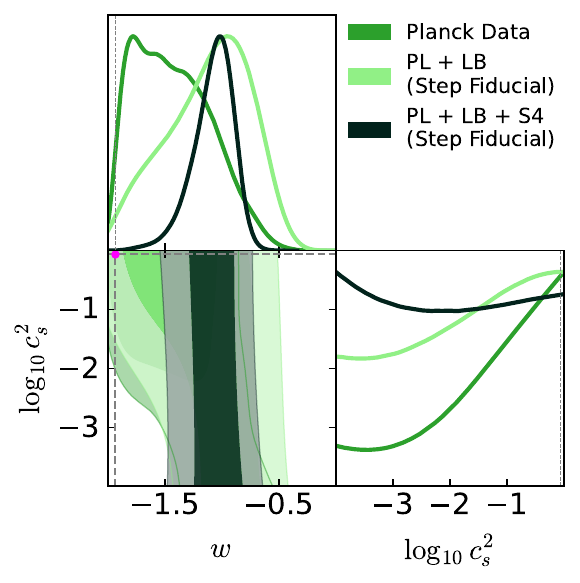}
     \end{subfigure}
     \caption{Constraints on the the kink [top left] and $w-c_s^2$ [top right] models, assuming a step fiducial. We mark the Planck best fit of each model with the small circles.}
     \label{fig:model_to_step}
\end{figure}

\begin{figure}[ht]
    \centering
    \includegraphics[width=\linewidth]{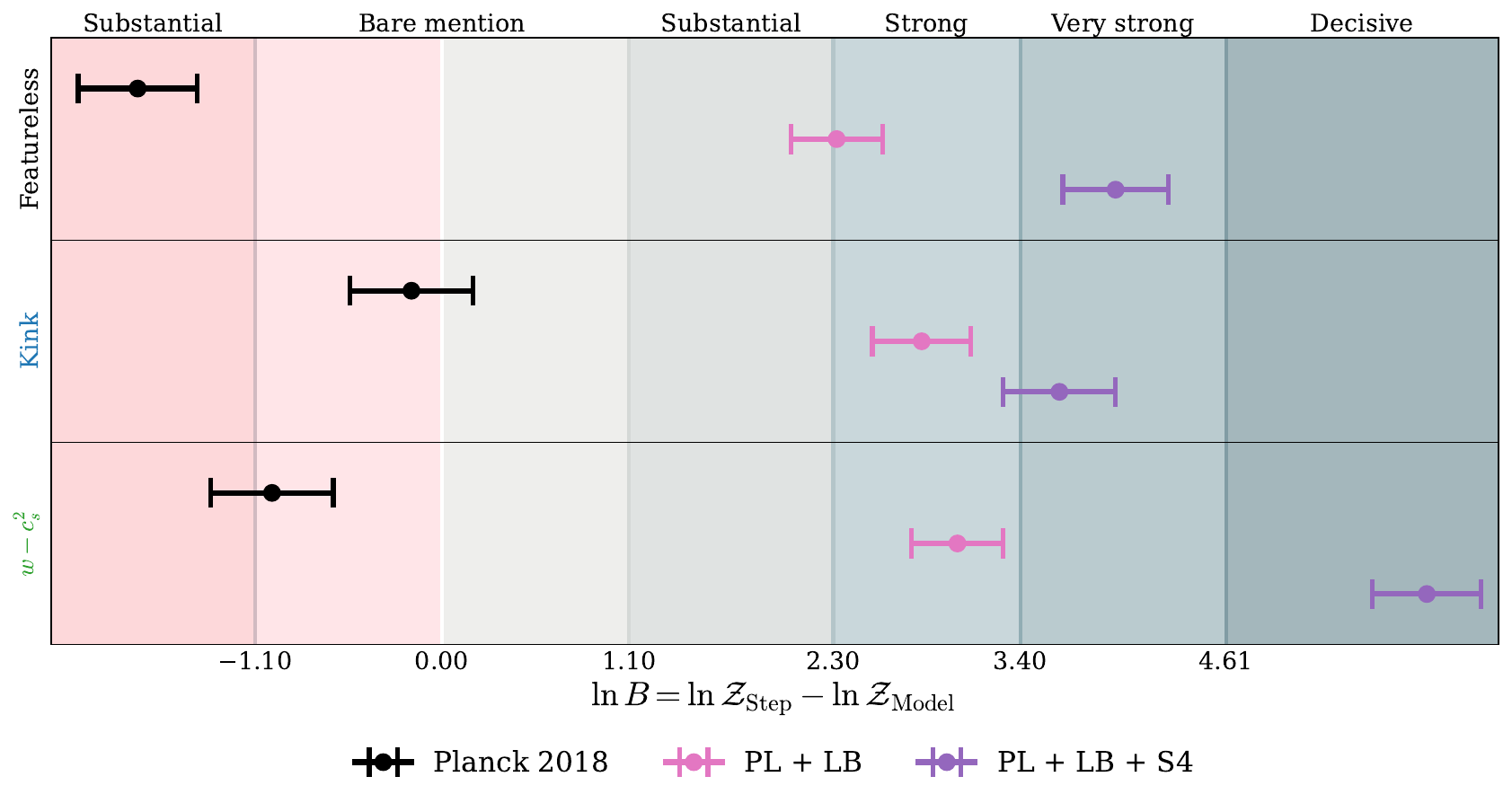}
    \caption{The projected Bayes factors obtained from our data and forecast analyses with respect to the step model.}
    \label{fig:model_to_step_bayes}
\end{figure}

\begin{figure}[ht]
    \centering
     \includegraphics[width=0.8\textwidth]{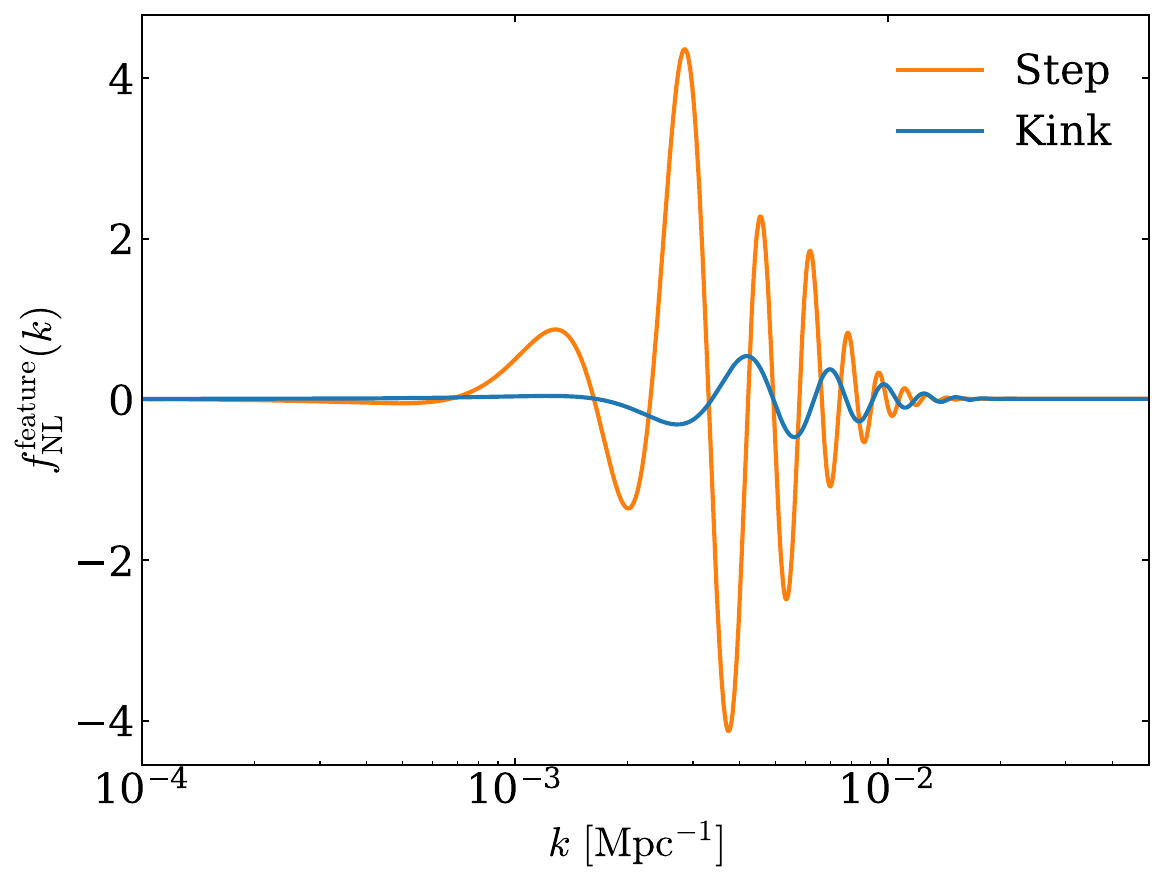}
     \caption{$f_{\rm NL}(k)$ in the equilateral limit for the best-fit step and kink model candidates.}
     \label{fig:fnl_step}
\end{figure}

Figure \ref{fig:model_to_step} shows the constraints on the effective parameters for the kink and $w-c_s^2$ models from the Planck data analysis and the forecast analyses, assuming a step fiducial, and Figure \ref{fig:model_to_step_bayes} shows the Bayes factors for these results. Compared to our analysis with real Planck data, the parameters of the kink model are more constrained. In particular, despite $\Delta \mathcal{P}_{\rm kink}/\mathcal{P}_0$ being consistent with $0$, we notice that its posterior peaks at a non-zero and negative value. The posterior of the parameter $N_0-N_*$ also displays a shift in the peak compared to the Planck data result, and smaller values of $\Delta N$ seem favored over larger ones. These are manifestations of the kink model trying to fit the large-scale fiducial dip. However, despite being slightly better than that for a featureless spectrum, the model is strongly disfavored over the step fiducial. Also, for the $w-c_s^2$ model, the results of the forecasts give similar results to the featureless fiducial --- $w$ is more tightly constrained around $w=-1$ and $c_s^2$ is unconstrained for the PL + LB + S4 forecast. Again, this is most likely due to better constraints on the cosmological parameters, especially from CMB-S4.

Considering the Bayes factors from this analysis, the featureless model is strongly disfavored with PL + LB. The simulated $E$-mode measurements, which are nearly cosmic-variance limited, allow for much better constraints of the large-scale dip-like feature. Such a signal cannot be reproduced by the featureless model. When including the simulated CMB-S4 measurements, the oscillations in the step fiducial at shorter scales become slightly more constrained, as seen from the improved constraint on $\Delta N$ in Figure \ref{fig:step_to_step}, which provides further evidence to rule out the featureless model. We conclude that if the true model of the universe includes a dip at large scales, it can be very strongly detected with these future experiments.

The kink model is strongly-to-very strongly disfavored compared to the step fiducial. Again, LiteBIRD provides much more precision of the large-scale $E$-mode anisotropies, which allows for better discrimination between the large-scale suppression (or amplification) of the kink model and large-scale dip of the step model. The addition of CMB-S4 provides a tighter constraint on the short-scale oscillations in the step model, as mentioned, and therefore provides slightly more evidence to rule out the kink model. Therefore, if the true model of the universe contains large-scale primordial features, there is strong-to-very strong evidence that future experiments will be able to distinguish between these different types of features.

Lastly, the $w-c_s^2$ model can be decisively ruled out if the true model of the universe contains a large-scale dip feature. First, considering the PL + LB forecast, the evidence is similar to that of the kink model forecast, as both models can result in large-scale suppression, which is not exhibited by the step fiducial. Therefore, when including the $EE$ and $TE$ spectra, which will be nearly cosmic-variance limited by LiteBIRD, both models can be strongly ruled out. However, with the addition of CMB-S4, since the shorter-scale oscillations in the step model are better constrained, the $w-c_s^2$ model is decisively ruled out as this model cannot capture such oscillations. In contrast, the kink model exhibits short-scale oscillations similar to the step model, while the only feature in the $w-c_s^2$ model is the suppression (or amplification) of power at $\ell\lesssim 10$. In other words, the kink model exhibits a more similar behavior at shorter scales to the step model, and thus adding CMB-S4 can only provide strong-to-very strong evidence against the kink model. Since these oscillations cannot be modeled by the $w-c_s^2$ model, we see decisive evidence to rule it out. Additionally, the $w-c_s^2$ model is further penalized by the addition of two useless parameters compared to the featureless model. Similar to the PL + LB + S4 forecast using the featureless fiducial, $w$ is constrained around $-1$ and $c_s^2$ becomes a useless parameter, so the inclusion of these two parameters does not provide any new information. The same can be said for the PL + LB + S4 step fiducial forecast, as $w$ is again constrained around a cosmological constant and $c_s^2$ becomes irrelevant. These results, combined with the short-scale oscillations in the step model, add a large amount of evidence against the $w-c_s^2$ model. Therefore, if the true model of the universe contains primordial features, we predict that such 2-parameter  dark energy models will be ruled out as the source of the low-$\ell$ anomalies with future experiments. Note, however, that LSS data will be able to provide much more accurate tests of dark energy models beyond $\Lambda$CDM.

Primordial non-Gaussianities may help to distinguish between different large-scale primordial feature models (see \cite{Chen:2010xka} for a review), such as the step and kink models. To this end, we briefly discuss the primordial bispectra predicted by the best fit step and kink models. The bispectrum $B(k_1,k_2,k_3)$ is defined as the three-point correlation of the curvature perturbation $\zeta$: $\langle \zeta_{\vec{k}_1} \zeta_{\vec{k}_2}\zeta_{\vec{k}_3} \rangle=(2\pi)^3\delta^{(3)}(\vec{k}_1+\vec{k}_2+\vec{k}_3)B(k_1,k_2,k_3)$. In Figure \ref{fig:fnl_step}, we plot the dimensionless number $f_{\rm NL}(k)$ in the equilateral limit $k_1=k_2=k_3=k$,
\begin{equation}
    f_{\rm NL}^{\rm feature}(k)=\frac{10}{9}\frac{1}{(2\pi)^4}\frac{k^6B(k,k,k)}{\mathcal{P}_0(k)^2},
\end{equation}
for the step and kink best-fit candidates to the Planck data, where we calculated $B(k,k,k)$ and $\mathcal{P}_0(k)$ using \texttt{BINGO}. Note that, in this definition, we use the featureless power spectrum $\mathcal{P}_0(k)$ so that the features in $f_{\rm NL}(k)$ come directly from the bispectrum and are not introduced by the power spectrum. We can see that both the phases of the running and the amplitudes are different. However, it would be challenging to observationally constrain $f_{\rm NL}^{\rm feature}$ at such an order of magnitude. We refer to \cite{Liguori:2010hx,Fergusson:2014hya,Fergusson:2014tza,Planck:2019kim} for works and reviews on such data analyses.

\section{Tests of low-$\ell$ anomalies ignoring short-scale effects} \label{sec:templates}
In our forecast analysis, we assumed that the true model of the universe was either featureless or contained a primordial step. In addition to the large dip that fits the $\ell\simeq 20$ anomaly in the $TT$ power spectrum, the step model also exhibits shorter-scale oscillations up to $\ell\approx 250$, and, importantly, the first of these oscillations following the dip has an amplitude which is (smaller but) comparable to the dip. This characteristic of the step model makes it testable across several multipoles, and our forecasts result in strong evidence in favor of the model (with the step fiducial) or against it (with the featureless fiducial). The kink model exhibits similar short-scale oscillations. It is, however, desirable to understand if future large-scale experiments, such as LiteBIRD, have the power to reveal the origin of the dip at $\ell\simeq 20$ and the suppression at $\ell\lesssim 10$ in the $TT$ data without appealing to detection of short-scale effects of the chosen model. To investigate this question, we consider two simplified diagnostic templates separately modeling the dip and suppression with no subsequent oscillations. While such templates are not motivated by realistic physics to our knowledge\footnote{A burst of particle production can indeed produce a Gaussian bump feature in the primordial power spectrum \cite{Barnaby:2009mc,Barnaby:2009dd,Naik:2022mxn}, but not a dip. Such a bump is also followed by linear oscillations which are a signature  of sharp features, but the envelope is such that their amplitude is very suppressed relative to the main bump. We could only find~\cite{Rodrguez:2020hot} in the literature in which a model displays a lognormal dip. However, it is unclear why that model does not exhibit the linear oscillations associated with sharp features.}, we consider these templates in order to test LiteBIRD's capabilities to detect such features independent of short-scale effects, rather than to analyze or constrain the templates themselves.

\subsection{Overview of the diagnostic templates}
To model the $\ell \simeq 20$ dip, we consider a lognormal feature in the primordial power spectrum of the form
\begin{equation}
    \mathcal{P}(k) = A_s \left(\frac{k}{k_*}\right)^{n_s-1} \left[ 1+ \frac{\Delta\mathcal{P}}{\mathcal{P}_0} \exp\left(-\frac{\ln^2(k/k_0)}{2 \sigma^2}\right)\right],
\end{equation}
where $\Delta\mathcal{P}/\mathcal{P}_0$ is the maximum amplitude of the dip, $k_0$ is the dip center, and $\sigma$ is the feature width. Similarly, we model the $\ell \lesssim 10$ suppression using a primordial spectrum of the form 
\begin{equation}
    \mathcal{P}(k) = A_s \left(\frac{k}{k_*}\right)^{n_s-1} \left[ 1- \frac{1}{2}\frac{\Delta\mathcal{P}}{\mathcal{P}_0} \left( \tanh\left(\frac{\ln(k/k_0)}{\sigma}\right)-1\right)\right],
\end{equation}
where $\Delta\mathcal{P}/\mathcal{P}_0$ is the difference between the large-scale suppression and the featureless spectrum. For both of these templates, we define $A_s$ and $n_s$ at the pivot scale $k_*\equiv 0.05~{\rm Mpc}^{-1}$, and we relate $k_0$ to the effective parameter $N_0-N_*$ via $\ln (k/k_0)=\ln(k/k_*)-(N_0-N_*)$. The prior ranges on the template parameters are listed in Table \ref{tab:bf_template}. We use our Bayesian inference pipeline to constrain the template parameters using the Planck data. We then find the best-fit candidates to the Planck data. Finally, we perform a forecast analysis, assuming the best-fit candidate of each template as the fiducial.

\subsection{Results for the dip template}

\begin{figure}
    \centering
    \includegraphics[width=\textwidth]{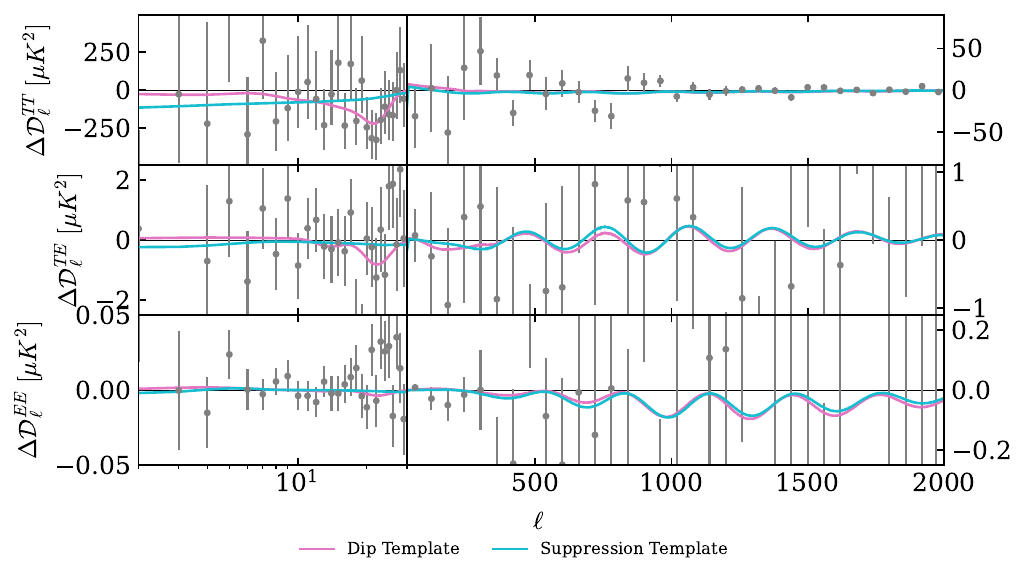}
    \caption{CMB power spectrum residuals of the best-fit candidates for the dip and suppression templates with respect to the best-fit featureless model. The residuals of the Planck measurements with respect to the best-fit featureless candidate are represented by the gray points. For a better visualization, we bin high-$\ell$ data using a binwidth $\Delta\ell=60$.}
    \label{fig:template_results1}
\end{figure}

Table \ref{tab:bf_template} lists the best-fit parameters to the Planck data for the diagnostic templates and the improvement in $\chi^2$ for these best-fit candidates over the featureless model. Figure \ref{fig:template_results1} shows the CMB power spectrum residuals for the best-fit candidates relative to the best-fit featureless candidate. Compared to the best-fit step model candidate (Table \ref{tab:models_bf} and Figure \ref{fig:bf}), the fit to the $\ell\simeq 20$ dip in the $TT$ spectrum is much improved by our dip template. The total $\Delta \chi^2$ and $\Delta \chi^2_{\rm lowT}$ for the best-fit dip template are larger than those for the best-fit step candidate, indicating a better overall fit to the data, primarily driven by the much better fit to low-$T$ data compared to the step model. This result demonstrates that the short-scale oscillations in the step model place more severe constraints on the model parameters, especially $\Delta \mathcal{P}/\mathcal{P}_0$, that prevent this model from fully capturing the $\ell\simeq 20$ dip.

\begin{table}[ht]
    \centering
    \begin{tabular}{|c|c|c|c|}
        \hline
        Parameters & Priors & Best-Fit Dip & Best-Fit Suppression \\ \hline
        $N_0-N_*$ & $[-5.0,-0.5]$ & -3.389021 & -3.032324 \\ \hline
        $\log_{10}\sigma$ & $[-2.0,\log_{\rm 10}(5)]$ & -0.9998020 & -1.693578 \\ \hline
        $\Delta\mathcal{P}/\mathcal{P}_0$ & $[-0.99,0.99]$ & -0.9563548 & -0.1307780 \\ \hline
        \end{tabular}
        
    \begin{tabular}{|c|c|c|c|c|c|}
        \hline
        Model & $\Delta \chi^2_{\rm total}$ & $\Delta \chi^2_{\rm low T}$ & $\Delta \chi^2_{\rm low E}$ & $\Delta \chi^2_{\rm EG20}$ & $\Delta \chi^2_{\rm prior}$ \\ \hline
        Dip & 7.068 & 6.579 & 0.165 & 0.186 & 0.138 \\ \hline 
        Suppression & 2.935 & 2.646 & -0.044 & 0.174 & 0.159 \\ \hline
    \end{tabular}

    \begin{tabular}{|c|c|c|c|}\hline
         & Planck 2018 & PL + LB & PL + LB + S4 \\ \hline
        Dip Fiducial & $-1.19\pm 0.35$ & $12.52\pm 0.28$ & $13.34\pm0.32$ \\ \hline
        Suppression Fiducial & $-1.53 \pm 0.35$ & $-1.39 \pm 0.26$ & $-1.59 \pm 0.30$ \\ \hline
    \end{tabular}
    \caption{[Top] Priors on the effective parameters and best-fit candidates to the Planck data for the dip and suppression templates. The priors on the cosmological parameters, $A_s$, and $n_s$ are the same as those listed in Table \ref{tab:priors}, and the nuisance parameters are unchanged from their default values. [Center] $\Delta \chi^2$ for the best-fit candidates with respect to the best-fit featureless candidate. [Bottom] Projected Bayes factors  $\ln B=\ln \mathcal{Z}_{\rm fiducial}-\mathcal{Z}_{\rm featureless}$ of the featureless model with respect to the dip and suppression fiducials obtained from our forecasts.}
    \label{tab:bf_template}
\end{table}

\begin{figure}[ht]
    \centering
    
    \begin{subfigure}[c]{0.49\textwidth}
         \centering
         \includegraphics[width=\textwidth]{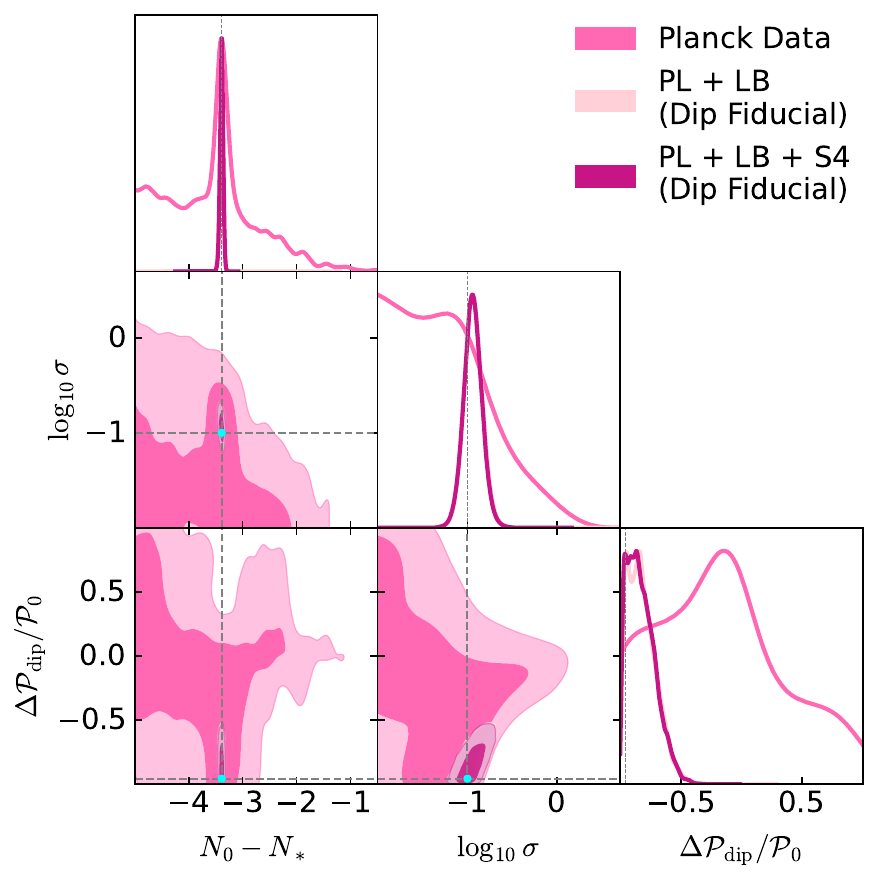}
     \end{subfigure}
     \hfill
     \begin{subfigure}[c]{0.49\textwidth}
         \centering
         \includegraphics[width=\textwidth]{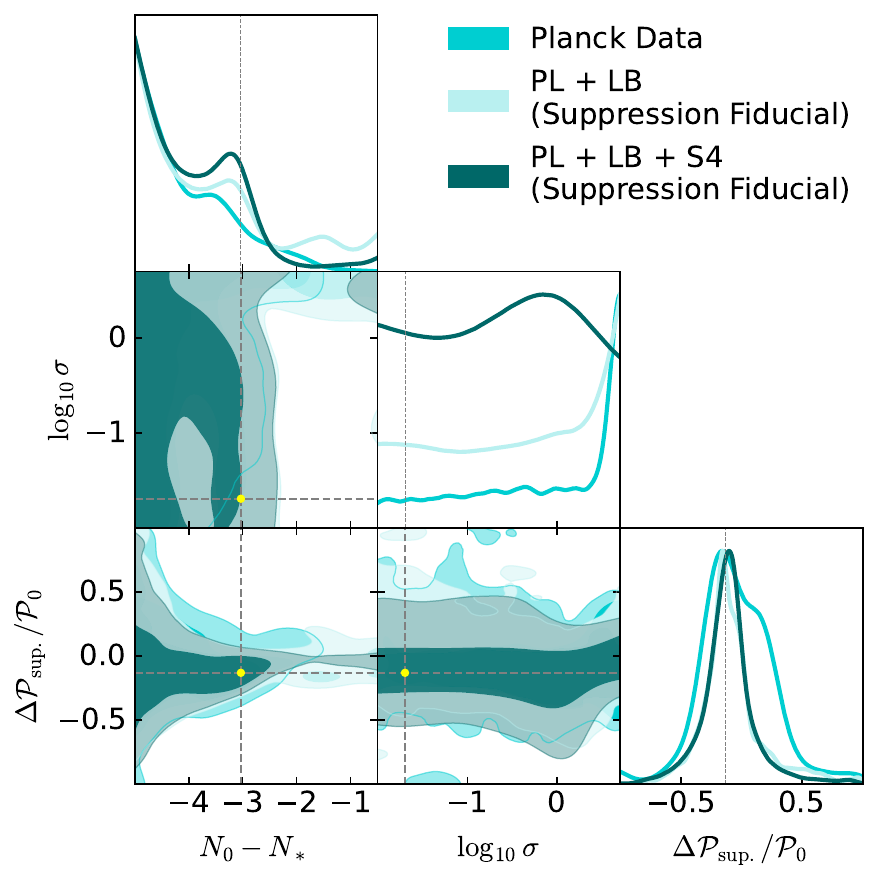}
     \end{subfigure}

     \begin{subfigure}[t]{0.49\textwidth}
         \centering
         \includegraphics[width=\textwidth]{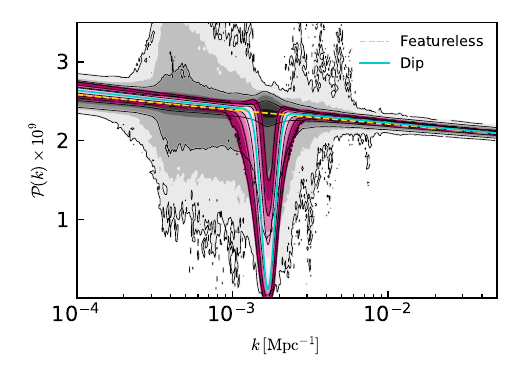}
     \end{subfigure}
     \hfill
     \begin{subfigure}[t]{0.49\textwidth}
         \centering
         \includegraphics[width=\textwidth]{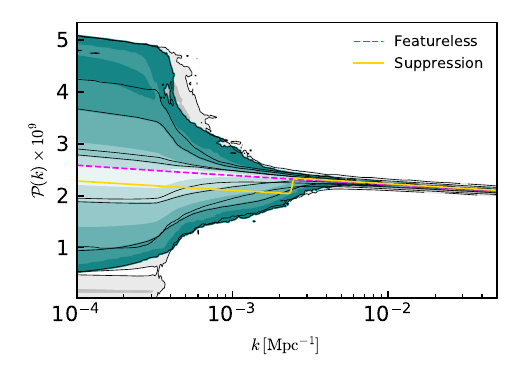}
     \end{subfigure}

     \caption{Results of the forecast analyses of the dip [left] and suppression [right] templates, assuming a dip and suppression fiducial, respectively. [Top] Constraints on the template parameters from our data and forecast analyses. [Bottom] Reconstructed primordial power spectra from the Planck data (gray colormap) and from our PL + LB + S4 forecasts (pink and teal colormaps). The best-fit featureless and template candidates are also plotted.
     }
    \label{fig:template_results2}
\end{figure}

We use this best-fit dip candidate as the fiducial in our forecast analysis to determine if a dip at large scales can be detected by upcoming CMB experiments, using the same methodology discussed in Sec. \ref{sec:forecast}. Constraints on the dip template parameters and reconstructions of the primordial power spectrum from our data and forecast analyses are shown in Figure \ref{fig:template_results2}. LiteBIRD will be able to place extremely tight constraints on the template parameters, while CMB-S4 will provide almost no improvement to these constraints since this template only affects large scales, which will be more precisely measured by LiteBIRD. Table \ref{tab:bf_template} lists the projected Bayes factors for the featureless model assuming the dip fiducial. With Planck data alone, there is substantial evidence in favor of the featureless model, primarily due to the large portion of the parameter space which is excluded by data---the fit of the model to data is, as discussed, indeed very good. However, in our forecasts, the dip feature can be decisively detected. We therefore conclude that the $\ell\simeq 20$ dip can indeed be detected by LiteBIRD if such a feature is exhibited in the true model of the universe, ignoring the oscillation after the dip and all short-scale effects.

\subsection{Results for the suppression template}

We now turn to discussing the data analysis for the suppression template, for which the results are given in Table \ref{tab:bf_template} and Figure \ref{fig:template_results1}. In contrast to the best-fit dip template, which yielded a larger amplitude and improvement in the fit compared to the best-fit step model, the amplitude of the best-fit suppression template is similar to that of the best-fit kink model candidate (see Table \ref{tab:models_bf}). This comparison demonstrates that a much smaller amplitude is needed to model the $\ell\lesssim 10$ suppression compared to the amplitude needed to model the $\ell \simeq 20$ dip. As a result, the short-scale oscillations in the kink model do not restrict the required suppression amplitude at large scales, as we saw for the step model. In fact, these small-amplitude short-scale oscillations allow for a slightly better fit to the data by the kink model compared to the suppression template at high-$\ell$ multipoles.

 We use this best-fit candidate for the suppression template as the fiducial in our forecast analysis to determine if such large-scale suppression can be detected by future observations. The projected Bayes factors for the featureless model from our suppression fiducial forecasts are listed in Table \ref{tab:bf_template}. From the Planck data alone, there is substantial evidence against the suppression template. In both of our forecasts, however, the suppression fiducial cannot be detected, and there is still substantial evidence in favor of the featureless model. 
 Figure \ref{fig:template_results2} shows the constraints on the suppression template parameters and reconstructions of the primordial power spectrum from our data and forecast analyses. The $N_0-N_*$ posteriors from both our data analysis and the forecasts show an increasing likelihood for small values, and the forecasts do not place tighter constraints around the best fit for this parameter. This is due to the large uncertainties at low-$\ell$ multipoles from cosmic variance combined with the relatively small amplitude of the suppression fiducial. Because of this small amplitude, the best-fit suppression template barely modifies the $EE$ spectrum at large scales, as seen in the CMB power spectrum residuals for the best fit. (In contrast, the best-fit dip template leaves a more significant deviation from featureless at low-$\ell$ multipoles in $EE$ that can be detected by LiteBIRD.) This means that the template parameters cannot be constrained. This effect is also demonstrated in the reconstructions of the primordial spectrum, which become completely unconstrained at large scales in both our data analysis and the PL+LB+S4 forecast. Furthermore, the feature width is unconstrained, especially in the PL+LB+S4 forecast, since the feature cannot be detected, and including CMB-S4 does not improve the precision at large scales. In summary, if the true model of the universe contains large-scale suppression with no short-scale effects, such a feature cannot be conclusively detected by LiteBIRD due to limited precision from cosmic variance and the relatively small amplitude of the suppression.

\section{Conclusions} \label{sec:concl}
In this study, we investigated the prospects of observing low-$\ell$ anomalies with future CMB temperature and polarization anisotropy measurements. Since cosmic variance in the $TT$ power spectrum has already been reached by Planck, we focused on the ability of LiteBIRD to provide nearly cosmic-variance limited measurements of $E$ modes at large scales. These predicted measurements, in combination with the sharp $EE$ transfer function, can provide tight constraints on models that exhibit large-scale deviations from the standard cosmological model.

We analyzed four different classes of models that generate three types of large-scale signals in the $TT$ power spectrum: a primordial dip at $\ell\simeq 20$, primordial $\ell<10$ suppression, late-time $\ell<10$ suppression, and $\ell<10$ amplification. In the first part of our analysis, we summarized the agreement of our chosen models with the Planck temperature and $E$-mode data using Bayesian inference. We found that all models, except the large-scale amplification, can improve the fit to data by fitting the low-$\ell$ anomalies, though none of them is statistically favored over the baseline model due to the extra number of parameters. Next, we performed a forecast of LiteBIRD and CMB-S4 measurements. We showed that, if the true model of the universe is the concordance model, we can strongly rule out a primordial large-scale dip and substantially-to-strongly rule out models predicting $\ell<10$ suppression. If the true model of the universe exhibits a dip at large scales, we predict that LiteBIRD and CMB-S4 would find strong evidence in favor of detection. Moreover, these experiments would be able to strongly rule out primordial large-scale suppression and decisively rule out late-time large-scale suppression, assuming the large-scale dip as the true universe model. Lastly, we tested whether LiteBIRD can detect large-scale features while neglecting short-scale modifications to the CMB power spectrum, such as those imposed by sharp primordial features. Our results showed that LiteBIRD can decisively detect the large-scale dip but cannot detect the large-scale suppression, without appealing to short-scale effects.

These findings demonstrate the ability of future CMB missions, namely LiteBIRD, to improve precision of measurements at large scales, allowing for definitive tests of the origin of the low-$\ell$ anomalies in the CMB temperature data. No other planned experiments will be capable of probing primordial fluctuations at such large scales, and only highly futuristic proposals --- such as measurements of the cosmic neutrino background~\cite{Zimmer:2024max} or the anisotropies of the stochastic gravitational wave background caused by the propagation of gravitational waves through density perturbations~\cite{Contaldi:2016koz,ValbusaDallArmi:2020ifo,Ricciardone:2021kel,Braglia:2021fxn,Galloni:2023pie} --- could potentially test primordial fluctuations at these scales. Furthermore, since primordial feature models typically display their signals across a wide range of scales, short-scale CMB missions, such as CMB-S4, can collaborate with LiteBIRD to pin down the more specific model of the universe by constraining the correlated short-scale oscillations.

In addition, data from LSS~\cite{Huang:2012mr,Hazra:2012vs,Chen:2016vvw,Ballardini:2016hpi,Palma:2017wxu,LHuillier:2017lgm,Ballardini:2017qwq,Beutler:2019ojk,Ballardini:2019tuc,Debono:2020emh,Li:2021jvz,Ballardini:2022wzu,Chandra:2022utq,Mergulhao:2023ukp,Euclid:2023shr} and 21cm hydrogen line surveys~\cite{Chen:2016zuu,Xu:2016kwz}, as well as the search for the stochastic gravitational wave background~\cite{Fumagalli:2020nvq,Braglia:2020taf,Fumagalli:2021dtd,Braglia:2024kpo} will provide new cosmological discoveries and exciting insights into fundamental physics within the coming decades at smaller scales.

\acknowledgments
We would like to thank Cora Dvorkin, John Kovac, and Avi Loeb for helpful discussions.
We thank Luis Anchordoqui and Ignatios Antoniadis for extensive discussions on the Dark Dimension model. MB and DKH
acknowledge travel support through the India-Italy
mobility program ’RELIC’ (INT/Italy/P-39/2022 (ER)).

\appendix
\section{Effective parameters for primordial feature models}
\label{app:eff_pars}
In this appendix, we compute the approximate relations between the model parameters and the effective parameters for the step and kink models. These models are defined by Eqs. (\ref{eq:step}) and (\ref{eq:kink}), respectively. We use the model parameters $A$, $\phi_0$, and $\Delta$, which we relate to the maximum power spectrum deviation from the featureless spectrum $\Delta \mathcal{P}/\mathcal{P}_0$, the spectral tilt at the feature location $n_s$, and the duration of the feature $\Delta N$, respectively. We also define the number of $e$-folds at which the feature occurs $N_0$. First, recall that the energy scale at the feature $\epsilon_0$ is related to $\phi_0$ and $n_s$ via Eq. (\ref{eq:eff_params1}). Since we keep $\epsilon_0$ fixed, $\phi_0$ is simply related to the effective parameter $n_s$.

Next, we consider the feature duration. The step and kink models introduce a feature width $\Delta \phi \sim C \Delta$, where $C$ is some constant of order unity. We approximate this constant using the full width at half maximum (FWHM) of $-\frac{1}{\Delta}{\rm sech}^2\left[\frac{\phi_0-\phi}{\Delta}\right]$, which we use as an ``effective width'' for its anti-derivative, $\tanh\left[\frac{\phi_0-\phi}{\Delta}\right]$. This leads to $\Delta \phi = 2\Delta~{\rm sech}^{-1}\left(\frac{1}{\sqrt{2}}\right) \approx 2\Delta$. We also note that the feature width is related to the number of $e$-folds the feature lasts, $\Delta N$, via $\Delta N=\Delta \phi/\sqrt{2\epsilon_0}$. This leads to the relation
\begin{equation}
    \Delta N \approx\frac{2\Delta}{\sqrt{2\epsilon_0}} ~.
    \label{eq:delta-N}
\end{equation}

We next consider the maximum deviation in the power spectrum from featureless. Note that the power spectrum amplitude is closely related to the inflaton velocity via $\mathcal{P}\approx H^4/(4\pi^2 \dot{\phi}^2)$. For the step model, we observe a bump/dip in the power spectrum resulting from the difference in the inflaton velocity at the feature location $\dot{\phi}_0$ compared to the attractor solution $\dot{\bar{\phi}}$. Since $H\sim$ constant, we have $\frac{1}{2}\dot{\phi}_0^2=\frac{1}{2}\dot{\bar{\phi}}^2+\Delta V$, where $\dot{\bar{\phi}}^2\approx\frac{2}{3}V_{\rm inf}\epsilon_0$ and $\Delta V=V_{\rm step}-V_0 \approx AV_{\rm inf}$ at $\phi=\phi_0$. Therefore, we can write the power spectrum deviation as
\begin{equation}
    \frac{\Delta \mathcal{P}_{\rm step}}{\mathcal{P}_0} =\frac{\mathcal{P}_{\rm step}-\mathcal{P}_0}{\mathcal{P}_0} \approx \left(1+\frac{3A}{\epsilon_0}\right)^{-1}-1 ,
    \label{eq:ps-step}
\end{equation}
where $\mathcal{P}_0$ is the featureless primordial spectrum.

For the kink model, the power spectrum amplitude before the feature differs from the amplitude after the feature. 
The asymptotic value of each amplitude is determined by the attractor solution, and therefore the value of  $dV/d\phi$. The change in the amplitude is then determined by the difference in $dV/d\phi$.
We therefore approximate the power spectrum deviation for the kink model,
\begin{equation}
    \frac{\Delta \mathcal{P}_{\rm kink}}{\mathcal{P}_0} =\frac{\mathcal{P}_{\rm kink}-\mathcal{P}_0}{\mathcal{P}_0}\approx \left(1+A\sqrt{\frac{2}{\epsilon_0}}\right)^{-1}-1 .
    \label{eq:ps-kink}
\end{equation}

\begin{figure}[ht]
    \centering
    \includegraphics[width=\textwidth]{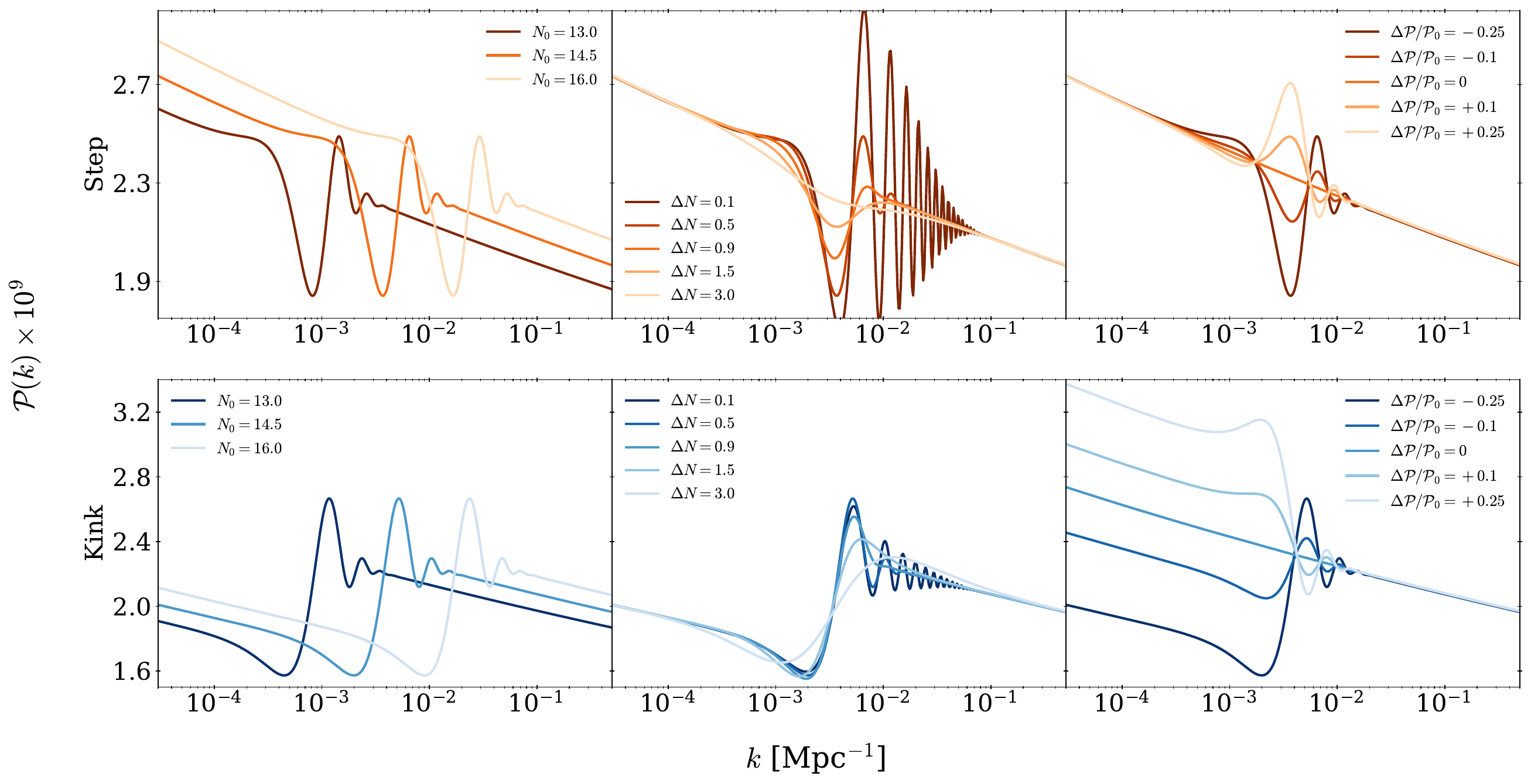}
    \caption{Examples of primordial power spectra with different effective parameters for the step model [top] and the kink model [bottom]. We fix $\ln (10^{10} A_s)=3.14$ and $n_s=0.966$. For both models, we start with the parameters $N_0=14.5$, $\Delta N=0.5$, and $\Delta \mathcal{P}/\mathcal{P}_0=-0.25$ and vary only one of these parameters at a time according to the values listed in the legends.}
    \label{fig:eff_params}
\end{figure}

Note that we define $\Delta \mathcal{P}/\mathcal{P}<0$ to correspond to a dip for the step model power spectrum and large-scale suppression for the kink model power spectrum. In terms of the model parameters, this corresponds to $A>0$, which means that the inflaton velocity after the feature (either the step or kink) is increased compared to the featureless case. Conversely, $\Delta \mathcal{P}/\mathcal{P}>0$ corresponds to a bump for the step model and large-scale amplification for the kink model in the power spectra (see Figure \ref{fig:eff_params}).

Lastly, we consider the feature location, parameterized by the number of $e$-folds after the start of inflation ($N_{\rm start}\equiv 0$) at which the feature occurs, $N_0$. Instead of determining $N_0$ with an explicit relation, we approximate the value using a shooting algorithm. We choose a starting value for the initial condition $\phi_i$, solve the equations of motion, and extract the feature location $N_0$ from the solution. We then adjust $\phi_i$ to obtain a desired value of $N_0$ within a precision of 0.01. We therefore obtain the effective parameter $N_0-N_*$, which can be related to the comoving scale of the feature $k_0$ by $N_0-N_*=\ln (k_0/k_*)$, where $k_*\equiv 0.05~{\rm Mpc}^{-1}$ is the scale which exits the horizon at $N_*=N_{\rm end}-50=18$ $e$-folds from the start of inflation. Note, that for these models, $A_s$ and $n_s$ are defined at the feature scale $k_0$ rather than at $k_*$. Examples of the primordial power spectra for both models with different effective parameters are shown in Figure \ref{fig:eff_params}.

\section{Discussion on the constraints on the dark dimension model}
\label{sec:DD-contour}     

\begin{figure}[ht]
    \centering
    \includegraphics[width=0.8\linewidth]{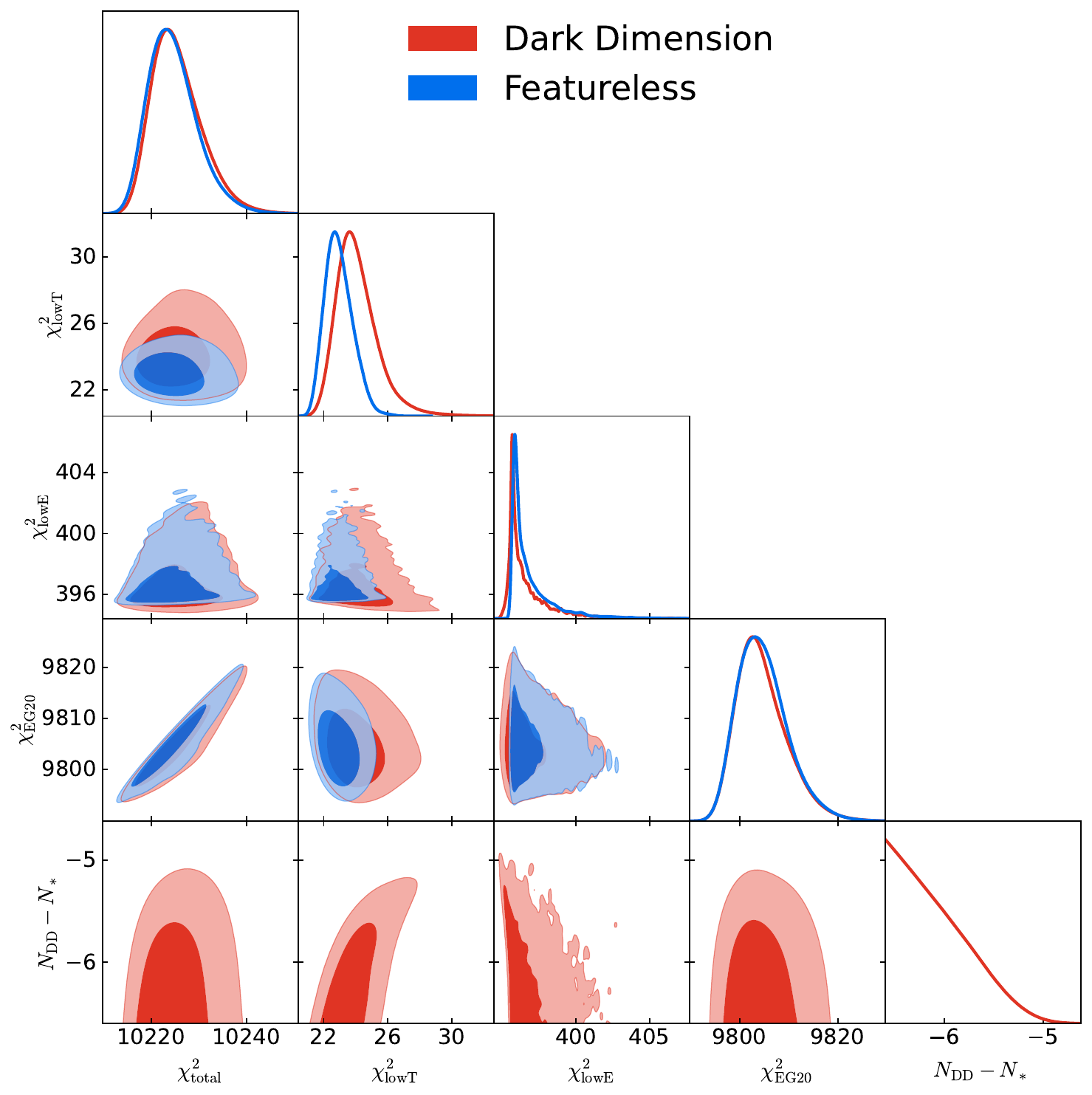}
    \caption{Marginalized posteriors for the total $\chi^2$, $\chi^2_{\rm lowT}$, $\chi^2_{\rm lowE}$, and $\chi^2_{\rm EG20}$ from the Planck analysis of the dark dimension model and the featureless model. The constraints on $N_{\rm DD}-N_*$ are also shown for the dark dimension model.}
    \label{fig:ddim_chi}
\end{figure}

In Figure~\ref{fig:ddim_chi}, we plot the effective total $\chi^2$ and the contribution of $\chi^2$ from the low multipole temperature ($\chi^2_{\rm lowT}$) and polarization ($\chi^2_{\rm lowE}$) data and the high-$\ell$ $TTTEEE$ ($\chi^2_{\rm EG20}$) data. Since amplification of the primordial power spectrum from the dark dimension model is limited to the largest scales only, we find the distribution of $\chi^2_{\rm EG20}$ (obtained from all the samples) to be nearly indistinguishable from the featureless baseline model. Since the Planck $EE$ data is not cosmic variance limited, polarization data also cannot distinguish between these models at large scales ($2\le\ell\le29$). Low-$\ell$ temperature data, however, does not favor the dark dimension model. The distribution of $\chi^2_{\rm lowT}$ significantly shifts towards a higher value compared to the featureless model at low multipoles, demonstrating that the temperature anisotropy data do not prefer amplification of power.

We find a strong upper bound on $N_{\rm DD}-N_*$. The correlation between $N_{\rm DD}-N_*$ and $\chi^2_{\rm lowT}$ indicates a worse fit to the data with amplification of power extending to larger multipoles. In the entire parameter space, the $\chi^2$ is worse than that of the featureless model, so that $N_{\rm DD}-N_*$ is pushed to small values at which the predicted deviation from the featureless spectrum becomes unobservable --- either on unobservable scales or within the large cosmic variance-limited uncertainties of the first few multipoles.
Larger $N_{\rm DD}-N_*$ values are ruled out by the data. 
Our analysis finds that $N_{\rm DD}-N_*<-5.38$ at 95\% confidence, which corresponds to a constraint on the angular multipoles of $\ell_{\rm DD}\lesssim 3.23$\footnote{To quote $N_{\rm DD}-N_*$ in terms of the multipole moment, we use the conversion $\ell_{\rm DD}=\eta_0\times k_{\rm DD}=\tau_0\times k_*e^{N_{\rm DD}-N_*}\sim14000 \, {\rm Mpc} \times k_*e^{N_{\rm DD}-N_*}$, where $k_*=0.05~{\rm Mpc}^{-1}$ and $\tau_0$ is the present day conformal time.}.
The results from these analyses indicate that the present CMB temperature and polarization data rejects the dark dimension model signature within the observable cosmological scales.

While the Bayes factor of this model is similar to those of the primordial feature models studied in this paper (i.e.~indicating substantial evidence against the models),
we should note the following difference. The dark dimension model always provides a worse fit to the data compared to the featureless baseline, even with an extra parameter. This includes the micron-sized dark dimension model highlighted in \cite{Anchordoqui:2023etp}, for which the change in $n_s$ occurs around $\ell_0\approx 10$, corresponding to $\ell_{DD}\approx 2$.
The other feature models studied in this paper, however, provide improved fits to the data compared to the baseline but are penalized for the extra degrees of freedom. With the upcoming cosmic variance-limited data from LiteBIRD, models providing a better fit to the data may get preference over the baseline model. 
On the other hand, the dark dimension model can be ruled in by the data only if future observations find temperature, polarization, and their cross-correlation anisotropy spectra to be higher in amplitude at large scales compared to the featureless model, a trend which is opposite to what has been observed in WMAP and Planck.
Another difference is that, if we increase the upper bound of the prior of the location parameter (i.e.~move these models to much shorter scales), the dark dimension model is ruled out in all of this extended parameter space, but the other feature models approach the featureless baseline model in the extended parameter space because these models only have effects in a finite range of scales. Thus, the dependence of the Bayes factors of these models on the choices of the prior range are different.

\bibliographystyle{JHEP}
\bibliography{main}

\end{document}